\documentclass[aps,amsmath,amssymb,twocolumn,nofootinbib,prr,superscriptaddress,longbibliography]{revtex4-2} 

\usepackage{hyperref} 
\usepackage{tabularx}
\usepackage[utf8]{inputenc}
\usepackage[T1]{fontenc}
\usepackage{graphicx}
\usepackage{bm}
\usepackage{epsfig,amsthm}
\usepackage{mathtools}
\usepackage{relsize}
\usepackage{accents}
\usepackage{wasysym}  
\usepackage{times}
\usepackage[mathscr]{euscript}
\usepackage[english]{babel} 

\usepackage[normalem]{ulem} 

\usepackage{amsmath,amsfonts,bm,thmtools}
\usepackage{mathptmx}
\usepackage{mathrsfs}
\DeclareMathAlphabet{\mathcal}{OMS}{cmsy}{m}{n}
\usepackage{xcolor}
\usepackage{multirow}

\definecolor{codegreen}{rgb}{0,0.6,0}
\definecolor{red}{rgb}{1,0,0}
\definecolor{nred}{rgb}{0.9,0.1,0.1}
\definecolor{nblack}{rgb}{0,0,0}
\definecolor{nblue}{rgb}{0.2,0.2,0.8}
\definecolor{ngreen}{rgb}{0.2,0.6,0.2}
\definecolor{ublue}{rgb}{0,0,0.5}
\definecolor{pur}{rgb}{0.75,0,0.75}
\definecolor{nngrn}{rgb}{0,0.5,0.5}
\definecolor{CitingColor}{rgb}{0,0.3,1}
\hypersetup{
	colorlinks=true,
	linkcolor=CitingColor,
	citecolor=CitingColor,
	urlcolor=CitingColor
}
\definecolor{codegreen}{rgb}{0,0.6,0}
\definecolor{red}{rgb}{1,0,0}
\hypersetup{
	colorlinks=true,
	linkcolor=CitingColor,
	citecolor=CitingColor,
	urlcolor=CitingColor
}

\newcommand{\ket}[1]{{|#1\rangle}}
\newcommand{\bra}[1]{{\langle#1|}}
\newcommand{\Tr}{{\rm Tr}}
\newcommand{\data}[1]{\vec{x}_#1}

\begin{document}

\title{Noise-enhanced quantum kernels on analog quantum computers for estimating the non-Markovianity from sparse temporal data}

\author{Hsiang-Wei Huang}
\affiliation{Department of Physics, National Cheng Kung University, 701401 Tainan, Taiwan}
\affiliation{Center for Quantum Frontiers of Research and Technology, NCKU, 701401 Tainan, Taiwan}

\author{Shen-Liang Yang}
\affiliation{Department of Physics, National Cheng Kung University, 701401 Tainan, Taiwan}
\affiliation{Center for Quantum Frontiers of Research and Technology, NCKU, 701401 Tainan, Taiwan}

\author{Chuan-Chi Huang}
\affiliation{Department of Engineering Science,National Cheng Kung University, 701401 Tainan , Taiwan.}
\affiliation{Physics Division, National Center for Theoretical Sciences, Taipei 106319, Taiwan}

\author{Yueh-Nan Chen}
\email{yuehnan@mail.ncku.edu.tw}
\affiliation{Department of Physics, National Cheng Kung University, 701401 Tainan, Taiwan}
\affiliation{Center for Quantum Frontiers of Research and Technology, NCKU, 701401 Tainan, Taiwan}
\affiliation{Physics Division, National Center for Theoretical Sciences, Taipei 106319, Taiwan}

\author{Hong-Bin Chen}
\email{hongbinchen@gs.ncku.edu.tw}
\affiliation{Center for Quantum Frontiers of Research and Technology, NCKU, 701401 Tainan, Taiwan}
\affiliation{Department of Engineering Science,National Cheng Kung University, 701401 Tainan , Taiwan.}
\affiliation{Physics Division, National Center for Theoretical Sciences, Taipei 106319, Taiwan}

\begin{abstract}
The quantum kernel method, a promising quantum machine learning algorithm, possesses substantial potential for demonstrating a quantum advantage. While most quantum kernels are constructed in the context of gate-based quantum circuits, inspired by the idea of analog quantum computing with Rydberg atoms, here we construct an analog quantum kernel and a hybrid quantum kernel, and demonstrate their competitiveness against other standard baselines in the physically motivated task of estimating the non-Markovianity from sparse temporal data. Crucially, we report a counterintuitive phenomenon where the operational noise can enhance their performance when interatomic distances are appropriately tuned. Our analysis reveals that the noise can help the analog and hybrid quantum models keep the test MSE under control despite closely fitting the training data, yielding models with higher complexity and a training MSE lower than or comparable to their ideal counterparts. Our results not only reveal that the noise can counterintuitively benefit analog and hybrid quantum kernel methods, but also provide an efficient route to estimating the non-Markovianity with alleviated experimental demands.

\end{abstract}

\flushbottom
\maketitle

\section{Introduction}

Along with the rapid growth in computational power, machine learning has emerged as one of the most powerful approaches for solving complex problems, and its success has also considerably influenced the field of quantum physics~\cite{carleo_mach_lear_appl_rmp_2019,mehta_mach_lear_appl_phys_rep_2019,
karniadakis_mach_lear_appl_nrp_2021,krenn_mach_lear_appl_pra_2023,hongming_mach_lear_stee_ms_cp_2024,hongbin_deep_learning_nonclassicality_qst_2025}. In parallel, quantum computation is also an emerging technology with the potential to surpass conventional computers~\cite{google_q_super_nature_2019,wright_benchmark_ionq_nc_2019,ibm_q_super_nature_2023,quantinuum_super_prx_2023,yunhua_process_n_cla_nvc_prr_2023}. Spurred by these advancements in both disciplines, there is growing interest in leveraging the unique features of quantum computation to demonstrate practical quantum advantage in machine learning algorithms~\cite{Biamonte_2017, Ciliberto_2018, Cerezo_2022}.
In recent years, various quantum machine learning algorithms~\cite{Schuld_2019,Havl_ek_2019,mitarai_qnn_pra_2018,edward_qnn_arxiv_2018,zoufal_qgan_npjqi_2019,fujii_q_res_comp_2020,Innocenti2023} have been proposed and have shown the potential to outperform their classical counterparts. Among these algorithms, the quantum kernel method is widely adopted and is arguably the most promising approach for realizing such quantum advantage~\cite{park_2020,Liu_2021,Huang_2021}.

Inspired by its classical counterpart, the quantum kernel method employs quantum feature maps to encode the data into high-dimensional quantum states. The similarities between data are then quantified using inner products in the Hilbert space, which play a crucial role in distinguishing samples in the original dataset. Beyond serving as a standalone algorithm, the quantum kernel method is a foundational component of supervised quantum machine learning~\cite{Schuld_2021_book}, and has accordingly attracted considerable attention. To gain deeper insights into quantum kernels, their expressivity and trainability have been studied within theoretical frameworks~\cite{Du_2020,Kubler_2021,Thanasilp_2024}, providing heuristic guidance for designing useful quantum kernels. On the application side, the quantum kernel method has been employed across a broad range of problems in quantum physics~\cite{Peters_2021,wu_qsvm_lhc_prr_2021,Sancho_2022,Wu_2023,Paine_2023,tancara_qsvm_non-mark_pra_2023,tsai_qsvm_steering_ms_njp_2025, Elkhouly_2025}, and its compatibility with quantum circuits has further enabled experimental demonstrations on various quantum devices~\cite{Peters_2021,wu_qsvm_lhc_prr_2021,Kusumoto2021, Bartkiewicz2020, Glick_2024}.

With the rapid development of quantum technology, gate-based quantum circuits have become the standard framework for expressing quantum algorithms. Various quantum feature maps, such as hardware-efficient ansatz (HEA)~\cite{Kandala2017}, angle encoding~\cite{angle_encoding}, dense angle encoding~\cite{dense_angle_encoding}, and ZZ feature map~\cite{Havl_ek_2019}, have been proposed in the context of gate-based quantum circuits and can be employed in the quantum kernel method. However, near-term quantum computers are still in the noisy intermediate-scale quantum (NISQ) era~\cite{Preskill_2018}, where substantial noise constitutes the primary challenge for quantum computation. This imposes a severe limitation on near-term applications. The output of a noisy quantum circuit is only reliable if its depth is sufficiently shallow, which in turn limits the quantum circuit complexity that can be realized in practice.

In contrast to gate-based digital quantum computation, an alternative paradigm is analog quantum computation, which solves computational tasks through controllable Hamiltonian evolution~\cite{ feynman1982simulating,Kendon2010,johnson2014quantum,Henriet2020,Jonathan_2023}. Classically intractable problems can be efficiently resolved by mapping them into the encoded Hamiltonian~\cite{S.Ebadi2022, Nguyen_2023,Jeong2023}. Building on this principle, a growing number of quantum algorithms have been implemented based on the principle of analog quantum computation~\cite{Martin_2020, Garcia-Molina2024, Martin2023, Lu_2025, Parra_2020}. A key advantage of this framework is that by replacing deep circuit blocks with continuous quantum evolution, these implementations substantially reduce the circuit depth. Spurred by these realizations, the concept of the quantum evolution kernel (QEK) has been proposed~\cite{Noori2020, Henry2021}, and various quantum feature maps based on this concept have been proposed~\cite{Yang_2023, Henry2021, Mehdi2026, Ayana2025, Noori2020, Albrecht2023}.

Building on this foundation, we construct quantum kernels in the context of analog quantum computing with Rydberg atoms. Rather than evaluating data similarity in classical space via renormalized feature vectors constructed from a sequence of quantum measurements~\cite {Henry2021, Noori2020}, we construct two possible architectures where similarity is estimated directly in the Hilbert space. The first, termed the analog quantum kernel, encodes data into the local energy difference of a Rydberg atom system to implement a feature map. The second, termed the hybrid quantum kernel, adopts the digital-analog ansatz structure~\cite{Lu_2025}, which combines the flexibility of gate-based digital quantum computation with the noise robustness inherent to analog evolution~\cite{Lu_2025}.

To evaluate the practical utility of these quantum kernels, we apply them to the physically motivated application of estimating the non-Markovianity from sparse temporal data. Non-Markovianity is a crucial property of open quantum system dynamics, arising from the inevitable interactions between a quantum system and its ambient environment~\cite{breuer_textbook,weiss_textbook,breuer_non_mark_review_rmp_2016,ines_non_mark_review_rmp_2017,hongbin_3sbm_scirep_2015}. Its underlying mechanism can be understood from the aspect of information exchange between them. Specifically, if a dynamical process leads to a monotonic outflow of information, the process is considered Markovian; conversely, any backflow of information serves as a distinct indicator of non-Markovianity. In order to witness the backflow of information, many approaches of quantification and characterization have been proposed~\cite{plenio_non_mark_review_rep_prog_phys_2014,BLP_measure_prl_2009,RHP_measure_prl_2010,LFS_measure_pra_2012,LPP_measure_pra_2013,BCM_measure_sci_rep_2014,francesco_prl_2014,chruscinski_k_divi_prl_2014,hongbin_k_divi_pra_2015,shinlian_tsw_measure_prl_2016,hongbin_n_mark_pra_2017}. Most of these methods hinge on monitoring the time evolution of certain quantities of interest and consider their revivals a signature of non-Markovianity. Therefore, to accurately capture their profiles, experimental realizations of these quantifiers typically demand the technique of quantum process tomography with sufficiently fine time resolution, thereby requiring substantial experimental resources~\cite{BLP_measure_exp_nat_phys_2011}, particularly when dealing with complex dynamics.

Aiming to alleviate the stringent experimental demands, we employ the kernel-based quantum machine learning algorithms to estimate the non-Markovianity from sparse temporal data. Additionally, we examine their performance in the presence of both operational and depolarizing noise. In the ideal setting, our numerical results show that the quantum models estimate the non-Markovianity with high accuracy, providing a highly efficient approach. Crucially, although noise is conventionally regarded as detrimental to performance, our numerical results reveal that these noisy quantum models can counterintuitively outperform their ideal counterparts when the interatomic distance falls within an appropriate range. Through quantitative analysis of these quantum models, we find that the noise can help the analog and hybrid quantum models keep test MSE under control despite closely fitting the training data. Consequently, these noisy quantum models exhibit higher model complexity and a training MSE lower than or comparable to that of their ideal counterparts. Additionally, since the noise lifts the near-zero tail of the ideal kernel eigenspectrum, we suggest that the noise induces a phenomenon analogous to tempered overfitting in the regime where the enhancement is observed. Ultimately, our results not only reveal that the noise can counterintuitively benefit analog and hybrid quantum kernel methods, but also provide an efficient route to estimating the non-Markovianity with alleviated experimental demands.

\section{Support vector regressor and kernel function}

We begin by briefly reviewing the principle of support vector machines (SVMs), which is one of the most widely used supervised learning algorithms.
Given a training dataset $\{(\vec{x}_j,y_j)\}$ of size $N$, comprising the features $\vec{x}_j\in\mathbb{R}^d$ and the corresponding labels $y_j\in\mathbb{R}$, an SVM is trained to capture the relationship between $\vec{x}_j$ and $y_j$, so that it can predict $y$ for previously unseen data $\vec{x}$. The basic model used in SVMs is a linear function $f(\vec{x}_j) = \vec{w}\cdot\vec{x}_j+b$ with trainable parameters $\vec{w}\in\mathbb{R}^d$ and $b\in\mathbb{R}$. During the training procedure, the trainable parameters are iteratively updated such that the deviations of the predictions $f(\vec{x}_j)$ from the ground-truth (GT) labels $y_j$ are no more than a desired error tolerance $\epsilon$, i.e., $|f(\vec{x}_j)-y_j|<\epsilon$. Several algorithms have been developed to solve this regression problem~\cite{Vijayakumar_1999, vapnik2000nature,schulz2018tutorial,hoerl1970,vovk2013kernel}.

Here we employ the support vector regression (SVR)~\cite{Vijayakumar_1999,vapnik2000nature} algorithm to solve the optimization problem. The SVR optimizes the trainable parameters $\vec{w}$ and $b$ by fitting a tube of width $\epsilon$ to the dataset. The hyperparameter $\epsilon$ defines the error tolerance by waiving the penalty for data samples residing within the $\epsilon$-tube. In contrast, samples falling outside the $\epsilon$-tube are penalized based on the deviation from the GT values. Therefore, the optimization problem is formulated as
\begin{equation}
\begin{aligned}
\text{minimize} \quad & \frac{1}{2} \Vert\vec{w}\Vert^2 + C\sum^N_{j=1} \left(\xi_j +\xi^*_j\right)  & \\
\text{subject to} \quad & \left\{\begin{aligned}
&y_j - (\vec{w}\cdot\vec{x}_j+b) \leq \epsilon+\xi_j\\
& (\vec{w}\cdot\vec{x}_j+b) - y_j \leq \epsilon+\xi_j^*\\
& \xi_j,~\xi_j^* \geq 0
\end{aligned}\right.,
\quad \forall~j
\end{aligned}
\label{eq_optimization_problem}
\end{equation}
The objective function penalizes incorrect predictions by the slack variables $\xi_j$ and $\xi^*_j\in\mathbb{R}$, and controls the balance between model complexity and training error by a hyperparameter $C > 0$. To solve the optimization problem~(\ref{eq_optimization_problem}), it is typically recast into its dual form~\cite{vapnik2000nature}
\begin{equation}
\begin{aligned}
\text{maximize} \quad & -\frac{1}{2} \sum_{j=1}^{N} \sum_{k=1}^{N} (\alpha_j - \alpha_j^*)(\alpha_k - \alpha_k^*) \langle\vec{x}_k,\vec{x}_j\rangle 
\\
&+ \sum_{j=1}^{N} (\alpha_j - \alpha_j^*) y_j - \epsilon \sum_{j=1}^{N} (\alpha_j + \alpha_j^*) \\
\text{subject to} \quad & \left\{\begin{aligned}
&\sum_{j=1}^{N} (\alpha_j - \alpha_j^*) = 0 \\
& 0 \leq \alpha_j,\alpha_j^* \leq C
\end{aligned}
\right.,
\quad \forall~j
\end{aligned}
\label{eq_optimization_dual}
\end{equation}
by further introducing the Lagrange multipliers $\alpha_j,\alpha_j^*\in\mathbb{R}$. Then the dual problem~(\ref{eq_optimization_dual}) can be efficiently solved by convex optimization techniques. Here, we denote the inner product as $\langle\vec{x}_k,\vec{x}_j\rangle=\vec{x}_k\cdot\vec{x}_j$ for convenience. At the optimum, the weight vector is given by $\vec{w}=\sum_{j=1}^N(\alpha_j - \alpha_j^*)\vec{x}_j$, yielding the desired decision function
\begin{equation}
f(\vec{x}) = \sum_{j=1}^{N} (\alpha_j - \alpha_j^*)\langle\vec{x},\vec{x}_j\rangle + b
\label{eq_decision_func_linear}
\end{equation}
for predicting a new sample $\vec{x}$.

The decision function~(\ref{eq_decision_func_linear}) is manifestly linear and its dependency on the similarity between data samples is captured by the inner product $\langle\vec{x}_k,\vec{x}_j\rangle$. In many practical problems, however, complicated datasets are generically nonlinearly separable, so the linear decision function~(\ref{eq_decision_func_linear}) cannot capture the nonlinear structure. Such datasets can often be made linearly separable by mapping them into an appropriate higher-dimensional feature space. This motivates the introduction of a feature map $\Phi(\vec{x}_j)$, which maps the original samples into a certain higher-dimensional feature space, and their similarity is quantified according to the kernel function $\langle \Phi(\vec{x}_k),\Phi(\vec{x}_j)\rangle$.

A further difficulty is that the feature map $\Phi$ and the associated feature space are generically computationally intractable or difficult to construct explicitly. This obstacle is circumvented by introducing the kernel function~\cite{aronszajn1950} $\kappa(\vec{x}_k,\vec{x}_j)=\langle \Phi(\vec{x}_k),\Phi(\vec{x}_j)\rangle$. Thanks to the representer theorem~\cite{representer_th}, the decision function $f(\vec{x})$ of an SVR can be expressed in terms of the kernel function as
\begin{equation}
f(\vec{x}) = \sum_{j=1}^{N} (\alpha_j - \alpha_j^*)\kappa(\vec{x},\vec{x}_j)+ b.
\label{eq_decision_function_kernel}
\end{equation}
The primary advantage of the kernel function is that it is efficiently computable without the need to explicitly construct $\Phi$. Once the kernel function is specified, the corresponding SVM can be trained. Due to its importance, various powerful kernel functions, such as the rational kernel~\cite{cortes_2004}, the convolutional kernel~\cite{haussler1999}, and the Gaussian radial basis function (RBF) kernel~\cite{rbf}, have been extensively studied in the literature. Spurred by the success of classical machine learning, attempts have been made to design the kernel function on quantum computers~\cite{Havl_ek_2019,Schuld_2019,Schuld_2021_book}, giving rise to the quantum kernel method.

\section{Quantum feature map and quantum kernel method}

Analogous to its classical counterpart, a quantum computer processes classical data with a quantum feature map $\Phi_\mathrm{Q}:\vec{x}_j\mapsto\ket{\psi_j}$ to encode data into high-dimensional quantum states. Due to the exponential growth of the Hilbert space with the number of involved qubits and the unique properties of entanglement, the quantum feature maps are considered highly effective at enhancing data expressivity compared to their classical counterpart~\cite{Du_2020}. Moreover, it has been shown that the quantum kernel method can outperform classical approaches in specifically tailored learning tasks~\cite{Liu_2021}.

\begin{figure*}[ht]
    \centering
    \includegraphics[width=\textwidth]{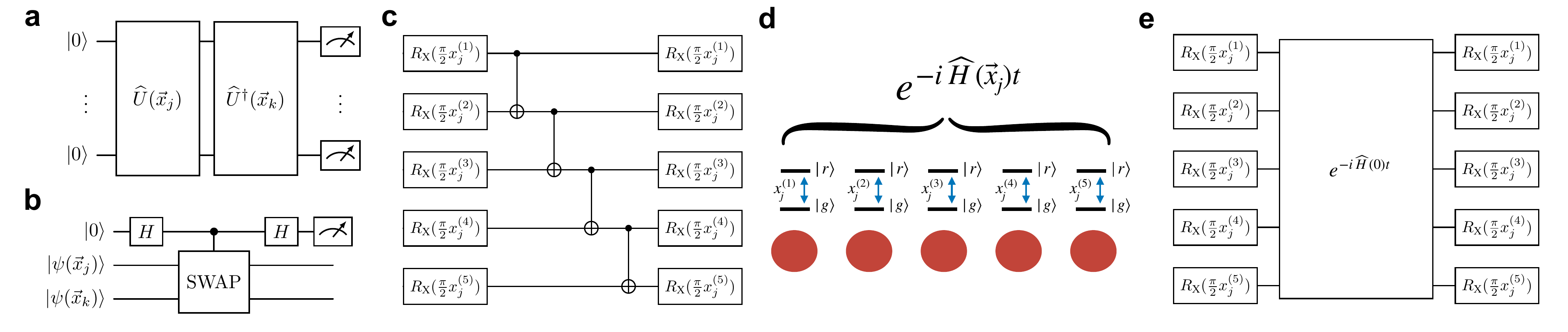}
    \caption{\textbf{Schematic illustrations of quantum kernel function and quantum feature maps encoding feature $\vec{x}_j~$.} \textbf{a} The quantum circuit can be used to estimate the quantum kernel function $\kappa_\mathrm{Q}(\vec{x}_k,\vec{x}_j)$ in Eq.~(\ref{eq_quantum_kernel}) by measuring the probability of the measurement outcome $\ket{0^{\otimes n}}$. \textbf{b} The SWAP test protocol can be employed for estimating the quantum kernel function when the adjoint of the quantum feature map is physically infeasible. \textbf{c} The HEA used here consists of two encoding layers sandwiching an entangling layer. The feature $\vec{x}_j$ is encoded into the single-qubit gates, and the sequence of CNOT gates will generate inter-qubit entanglement. \textbf{d} In an analog quantum feature map, the feature $\vec{x}_j$ is encoded into the Hamiltonian operator $\widehat{H}(\vec{x}_j)$, which governs the autonomous quantum evolution of the computing agent. Here we consider an array of atoms arranged in a straight line at the nearest interatomic distance $a$. The feature is encoded into the energy-level spacings. \textbf{e} We also consider a hybrid type by replacing the sequence of CNOT gates in the HEA with an analog layer governed by a degenerate Hamiltonian $\widehat{H}(0)$.}
    \label{fig:qsvm}
\end{figure*}

The quantum feature map $\Phi_\mathrm{Q}$ is realized by a quantum operation $\widehat{U}(\vec{x}_j)$ parameterized by the input features. The resulting data-dependent quantum states are given by $\ket{\psi(\vec{x}_j)}=\widehat{U}(\vec{x}_j)\ket{0^{\otimes n}}$. Based on the concept of the kernel function, the quantum kernel function is defined in a similar manner~\cite{Havl_ek_2019,Schuld_2019,Schuld_2021_book} as
\begin{equation}
\kappa_\mathrm{Q}(\vec{x}_k,\vec{x}_j)
=\lvert\langle\psi(\vec{x}_k) | \psi(\vec{x}_j)\rangle\rvert^2
=\lvert\bra{0^{\otimes n}}\widehat{U}^\dagger(\vec{x}_k)\widehat{U}(\vec{x}_j)\ket{0^{\otimes n}}\rvert^2,
\label{eq_quantum_kernel}
\end{equation}
which quantifies the similarity between two states $\ket{\psi(\vec{x}_j)}$ and $\ket{\psi(\vec{x}_k)}$. Experimentally, the quantum kernel function can be efficiently estimated by measuring the probability of the measurement outcome $\ket{0^{\otimes n}}$ on the quantum state $\widehat{U}^\dagger(\vec{x}_k)\widehat{U}(\vec{x}_j)\ket{0^{\otimes n}}$, as shown in Fig.~\ref{fig:qsvm}\textbf{a}. If the adjoint of the quantum feature map is physically infeasible, the SWAP test protocol~\cite{Buhrman2001} shown in Fig.~\ref{fig:qsvm}\textbf{b} can instead be employed to evaluate the quantum kernel function with the aid of an additional ancilla qubit. The value of the quantum kernel function $\kappa_\mathrm{Q}(\vec{x}_k,\vec{x}_j)
=2p_{\ket{0}}-1$ is then estimated from the probability $p_{\ket{0}}$ of obtaining the outcome $\ket{0}$ upon measuring the ancilla qubit.

Since the quantum kernel function is the building block of Eq.~\eqref{eq_decision_function_kernel}, the specific architecture of $\widehat{U}(\vec{x}_j)$ strongly influences the performance and properties of the decision function. Nevertheless, the optimal design of $\widehat{U}(\vec{x}_j)$ for a given learning task remains an open question. Here, we focus on three particular types of quantum feature maps.

\subsection{Digital quantum kernel}

In most practical scenarios, the design of $\widehat{U}(\vec{x}_j)$ is mostly empirical, depending on the specific characteristics of the quantum devices that implement the circuit. One of the most widely adopted architectures for gate-based digital quantum computers is the hardware-efficient ansatz (HEA)~\cite{Kandala2017}, which exists in numerous variants. For our purpose here, we adopt the following structure:
\begin{equation}
\widehat{U}_\mathrm{D}(\vec{x}_j)=
\left[\bigotimes_{\mu=1}^d R_\text{X}(\frac{\pi}{2}x_j^{(\mu)})\right]
\left[\overrightarrow{\prod}_{\mu=1}^{d-1}\text{CNOT}(\mu,\mu+1)\right]
\left[\bigotimes_{\mu=1}^d R_\text{X}(\frac{\pi}{2}x_j^{(\mu)})\right].
\label{digtal encoding}
\end{equation}
It comprises two types of layers: the encoding layer and the entangling layer. For a given feature $\vec{x}_j=(x_j^{(1)},\ldots,x_j^{(d)})$, the former encodes each component $x_j^{(\mu)}$ into single-qubit rotation gates; while the latter generates entanglement by a sequence of CNOT gates, as shown in Fig.~\ref{fig:qsvm}\textbf{b}.

Since the CNOT gate is the primary source of hardware noise, we adopt a linear arrangement in the entangling layer, exclusively entangling adjacent qubits. This requires $d-1$ CNOT gates in an entangling layer. Alternative structures, such as circularly connected or fully connected arrangements, will substantially increase the number of CNOT gates, leading to increased gate noise during circuit execution.

\subsection{Analog quantum kernel}\label{sec:aqk}

In addition to the HEA implemented on gate-based quantum devices, here we consider an alternative quantum feature map constructed according to the principles of analog quantum computation. In this analog quantum kernel, the feature $\vec{x}_j$ is encoded into the Hamiltonian operator $\widehat{H}(\vec{x}_j)$ of the computing agent, which then evolves autonomously according to the quantum evolution operator
\begin{equation}
\widehat{U}_\mathrm{A}(\vec{x}_j)=\exp[-i\widehat{H}(\vec{x}_j) t]
\end{equation}
for a finite time period. Unlike a gate-based circuit, this kind of quantum feature map may not be readily inverted, which dictates the use of the SWAP test protocol discussed above.

A prominent candidate for implementing the analog quantum feature map is the Rydberg atom system~\cite{Adams_2020, Gallagher_1988}. In this system, the computing agent is a group of atoms, in which the atom is selectively excited via the transition $\ket{g}\leftrightarrow\ket{r}$ between the ground state $\ket{g}$ and a specific high-energy state (Rydberg state) $\ket{r}$. For a system consisting of $d$ atoms, the corresponding Hamiltonian is given by:
\begin{align}
\widehat{H}(\vec{x}_j) =&  \frac{\Delta}{2}\sum_{\mu=1}^d x_j^{(\mu)}\hat{\sigma}_z^{(\mu)}
+\frac{\Omega}{2}\sum_{\mu=1}^d \hat{\sigma}_x^{(\mu)}
\\
&+\sum_{\mu<\nu}V_{\mu\nu}\left(\ket{r}\bra{r}\right)_\mu\left(\ket{r}\bra{r}\right)_\nu,
\label{eq_rydberg_hamiltonain}
\end{align}
where $\hat{\sigma}_x^{(\mu)}$ and $\hat{\sigma}_z^{(\mu)}$ are the Pauli operators acting on the $\mu$th atom. The first term in Eq.~(\ref{eq_rydberg_hamiltonain}) represents the energy-level spacing between $\ket{g}$ and $\ket{r}$, into which the feature $\vec{x}_j$ is encoded. The second term describes the Rabi oscillation with Rabi frequency $\Omega$. In principle, it can be tuned individually for each atom. For simplicity, we set $\Omega=8\pi$ MHz for all atoms, and fix the ratio $\Delta/\Omega=0.5$. The evolution time is set to $t=2.5\times10^{-1}$ $\mu$s such that the evolution lasts for a Rabi cycle, i.e., $\Omega t=2\pi$.

Crucially, the third term in Eq.~(\ref{eq_rydberg_hamiltonain}) denotes the van der Waals interaction induced by the electric dipole~\cite{Gallagher_1988}. For two atoms at positions $\vec{r}^{(\mu)}$ and $\vec{r}^{(\nu)}$, the interaction strength scales as $V_{\mu\nu}\propto|\vec{r}^{(\mu)}-\vec{r}^{(\nu)}|^{-6}$. Therefore, the effects of this interaction are highly sensitive to the atomic arrangement. For instance, when two atoms are close enough to each other, the Rydberg blockade effect~\cite{Lukin_2001, Urban_2009} occurs, strongly suppressing the doubly excited state. This phenomenon occurs when the interatomic distance is less than a critical value, i.e., the blockade radius $R_\mathrm{b}$, at which $V_{\mu\nu}\approx\Omega$. For simplicity, we consider a linear configuration in which $d$ atoms are arranged in a chain with a nearest-neighbor distance $a$, as shown in Fig.~\ref{fig:qsvm}\textbf{c}.

\subsection{Hybrid quantum kernel}

Following the digital-analog ansatz proposed in Ref.~\cite{Lu_2025}, we also consider a hybrid type of quantum feature map, in which the sequence of CNOT gates in the HEA is replaced by an analog layer, as shown in Fig.~\ref{fig:qsvm}\textbf{d}. In our construction, the feature $\vec{x}_j$ is encoded into the single-qubit rotation gates through the encoding layers, while the analog layer is governed by a degenerate Hamiltonian $\widehat{H}(0)$. Accordingly, the hybrid quantum feature map is expressed as:
\begin{equation}
\widehat{U}_\mathrm{H}(\vec{x}_j)=
\left[\bigotimes_{\mu=1}^d R_\text{X}(\frac{\pi}{2}x_j^{(\mu)})\right]
\exp[-i \widehat{H}(0) t]
\left[\bigotimes_{\mu=1}^d R_\text{X}(\frac{\pi}{2}x_j^{(\mu)})\right].
\end{equation}

In this construction, interatomic entanglement is established via van der Waals interactions during autonomous evolution~\cite{bernien2017}, and the analog layer serves as an entangling layer with the capability to achieve maximal scrambling~\cite{Abd-Rabbou2026}. Importantly, because the sign of the van der Waals interaction is fixed by the choice of Rydberg state and cannot be dynamically reversed within a single execution, the analog and hybrid quantum kernels can be estimated only through the SWAP test protocol.

\subsection{Noise models}\label{subsec:Noise model}

Since current quantum devices are still in the NISQ era, their performance is significantly affected by inherent noise. We are therefore interested in how noise impacts the quantum models. 
Specifically, we consider the operational noise caused by the imperfect manipulation of gate parameters $\vec{\upsilon}$ of certain quantum gates and assume that the remaining noise can be approximated by a depolarizing map
\begin{equation}
    \Lambda_{\textrm{dpl}}(\rho) = (1-p)\rho + p\frac{1}{2^d}\widehat{I},
\end{equation}
which maps a quantum state $\rho$ to the maximally mixed state with probability $p$.

Considering that single-qubit gates exhibit lower noise levels in most quantum devices, we exclusively focus on the operational noise in the entangling layers, including the analog layer within both analog and hybrid kernels, and the CNOT-gate layer in the HEA. For the case of analog layer, the noisy parameters $\vec{\upsilon}$ are played by the tunable parameters in the Hamiltonian~(\ref{eq_rydberg_hamiltonain}), including
\begin{enumerate}
\item the noisy energy-level spacing $\tilde{\Delta} \sim \Delta+\mathcal{N}(0,0.1~\text{MHz})$,

\item the noisy Rabi frequency $\tilde{\Omega}\sim\Omega\cdot \mathcal{N}(1,0.01)$,

\item and the fluctuating atomic positions $\tilde{r}^{(\mu)}\stackrel{\text{iid}}{\sim} r^{(\mu)} + \mathcal{N}(0,0.1~\mu\text{m})$,
\end{enumerate}
where $\mathcal{N}(\mathrm{mean},~\mathrm{std.})$ is a Gaussian distribution characterized by mean and standard deviation. The relevant parameters are extracted from the real device~\cite{Jonathan_2023}.

Additionally, to model the operational noise in the CNOT gates, we first observe that a conventional CNOT gate, $\text{CNOT}(1,2)$, acting on the second qubit while conditioned on the first qubit, can be expressed as
\begin{equation}
    \text{CNOT}(1,2)= e^{-i\theta(\widehat{I}^{(1)} -\hat{\sigma}_z^{(1)})\otimes(\widehat{I}^{(2)} - \hat{\sigma}_x^{(2)})}
\end{equation}
with $\theta=\pi/4$. In this case, the noisy parameter $\upsilon$ is a single random variable played by $\theta$ satisfying the Gaussian distribution $\mathcal{N}(\pi/4, 0.035)$, resulting in an average CNOT gate fidelity of 99\%.

After establishing the noise models, we proceed to simulate the effects of the noise. Specifically, for each individual data sample $\vec{x}_j$ and the corresponding quantum feature map $\widehat{U}(\vec{x}_j)$, the effects of the noise can be simulated by sampling the noisy parameters $\vec{\upsilon}$ for $M$ times and constructing the corresponding noise ensemble $\left\{\ket{\psi(\vec{x}_j,\vec{\upsilon}_1)}, \ket{\psi(\vec{x}_j,\vec{\upsilon}_2)},\ldots,\ket{\psi(\vec{x}_j,\vec{\upsilon}_M)}\right\}$, where each single realization is given by $\ket{\psi(\vec{x}_j,\vec{\upsilon}_m)}=\widehat{U}(\vec{x}_j,\vec{\upsilon}_m)\ket{0^{\otimes n}}$. From this noise ensemble, we obtain the noisy data-encoded density matrix
\begin{equation}\label{eq:representation of noisy state}
\tilde{\rho}(\vec{x}_j)=\frac{1}{M}\sum_{m=1}^M \ket{\psi(\vec{x}_j,\vec{\upsilon}_m)}\bra{\psi(\vec{x}_j,\vec{\upsilon}_m)}
\end{equation}
by mixing the noise ensemble. After applying the depolarizing map on $\tilde{\rho}(\vec{x}_j)$, the resulting noisy kernel function is given by
\begin{equation}
\tilde{\kappa}_\mathrm{Q}(\vec{x}_k,\vec{x}_j)= (1-p)^2\Tr \left[\tilde{\rho}(\vec{x}_k) \tilde{\rho}(\vec{x}_j)\right] + \frac{p(2-p)}{2^d}.
\label{eq:noisy quantum kernel}
\end{equation}
The corresponding noisy kernel matrix $\widetilde{K}_\mathrm{Q}$ reads
\begin{equation}
    \widetilde{K}_\mathrm{Q} = (1-p)^2\widetilde{K}^{\mathrm{op}}_\mathrm{Q} + \frac{p(2-p)}{2^d}\widetilde{J},
\end{equation}
where $\widetilde{K}^{\mathrm{op}}_\mathrm{Q}=\sum_{k,j}\Tr \left[\tilde{\rho}(\vec{x}_k) \tilde{\rho}(\vec{x}_j)\right]\ket{k}\bra{j} $ is the noisy kernel matrix under the operational noise, and $\widetilde{J}$ is the matrix of ones. From the constraints on dual variables in Eq.~\eqref{eq_optimization_dual}, it follows that the resulting model trained by $\widetilde{K}_\mathrm{Q}$ differs from that trained by $\widetilde{K}^{\mathrm{op}}_\mathrm{Q}$ only by a scaling factor, as detailed in Methods. Accordingly, in the following discussion, we employ $\widetilde{K}^{\mathrm{op}}_\mathrm{Q}$ to train the noisy models.

\section{Estimation of non-Markovianity from sparse raw data}\label{subsec:intro2nm}

\begin{figure*}[!ht]
\includegraphics[width=\textwidth]{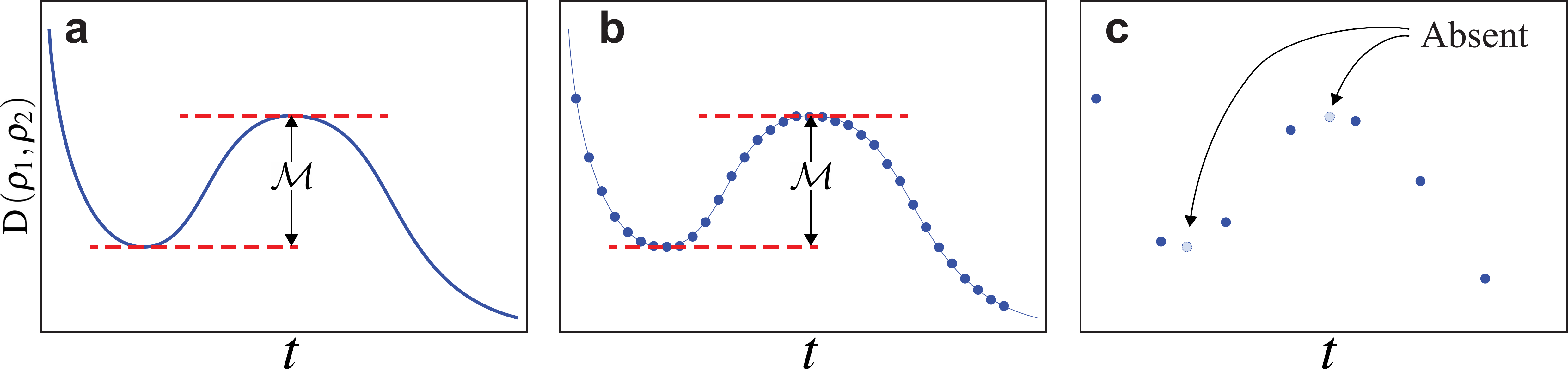}
\caption{\textbf{Quantification of the non-Markovianity and estimation from sparse temporal data.} \textbf{a} The underlying idea of the BLP non-Markovianity measure (\ref{eq_blp_measure}) is to pursue the maximal cumulative revival of the trace distance $\mathrm{D}\left(\rho_1(t),\rho_2(t)\right)$ of a certain optimal initial state pair. \textbf{b} In an experimental implementation of the BLP measure, to accurately identify the local extrema, a substantial amount of experimental raw data with sufficiently fine time resolution is required. \textbf{c} A coarse time resolution may prevent the precise identification of local extrema, leading to an unreliable estimation. Consequently, we explore how the non-Markovianity can be estimated from sparse data in the absence of precise local extrema by leveraging the power of machine learning approaches.}
\label{fig_nm_measure_illustration}
\end{figure*}

After presenting the theoretical framework of our approach, we apply all the aforementioned models to the problem of estimating the non-Markovianity from sparse temporal data. We begin by briefly reviewing the concept of the non-Markovianity measure and elucidating the challenges encountered in experimental implementations.

Due to the inevitable interactions between a quantum system and its ambient environment, the system dynamics typically exhibit incoherent dynamical behavior~\cite{breuer_textbook,weiss_textbook,breuer_non_mark_review_rmp_2016,ines_non_mark_review_rmp_2017,hongbin_3sbm_scirep_2015}. During these interactions, intricate correlations are established. However, subject to fluctuations within the vast environmental degrees of freedom, these correlations are typically fragile and transient, and their destruction constitutes the primary cause of decoherence. As a result, information is continuously exchanged between the system and its environment. If a dynamical process leads to a monotonic outflow of information, the process is considered Markovian; conversely, any backflow of information serves as a signature of non-Markovianity. Additionally, the direction of information flow has been mathematically associated with the divisibility of the dynamical map into completely positive subchannels. Based on these insights, extensive efforts have been devoted to the quantification and characterization of non-Markovianity~\cite{plenio_non_mark_review_rep_prog_phys_2014,BLP_measure_prl_2009,RHP_measure_prl_2010,LFS_measure_pra_2012,LPP_measure_pra_2013,BCM_measure_sci_rep_2014,francesco_prl_2014,chruscinski_k_divi_prl_2014,hongbin_k_divi_pra_2015,shinlian_tsw_measure_prl_2016,hongbin_n_mark_pra_2017}.

Among the various proposed quantifications, we specifically focus on the measure introduced by Breuer, Laine, and Piilo (BLP)~\cite{BLP_measure_prl_2009}. The BLP measure quantifies the information contained within a system by the trace distance $\mathrm{D}\left(\rho_1,\rho_2\right)=\Vert\rho_1-\rho_2\Vert_{1}/2$ between two quantum states, $\rho_{1}$ and $\rho_{2}$, where
$\Vert A\Vert_{1}=\Tr(\sqrt{A^\dagger A})$ denotes the trace norm of a matrix $A$. When the trace distance decreases monotonically, its time-varying rate $\dot{\mathrm{D}}\left(\rho_1(t),\rho_2(t)\right)=d \mathrm{D}\left(\rho_1(t),\rho_2(t)\right)/dt \leq 0$, implying that information is continuously flowing out of the system, which characterizes Markovian dynamics. Conversely, $\dot{\mathrm{D}}\left(\rho_1(t),\rho_2(t)\right)>0$ signifies a backflow of information from the environment into the system, which temporarily increases the distinguishability of the state pair and indicates the non-Markovian nature of the dynamics. Based on this idea, the BLP measure quantifies non-Markovianity according to
\begin{equation}
\mathcal{M}=\max_{\rho_1(0),\rho_2(0)}\int_{\dot{\mathrm{D}}>0}
\dot{\mathrm{D}}\left(\rho_1(t),\rho_2(t)\right) dt,
\label{eq_blp_measure}
\end{equation}
pursuing the cumulative revival of the trace distance maximized over all initial state pairs, as schematically shown in Fig.~\ref{fig_nm_measure_illustration}\textbf{a}. The BLP measure has been experimentally implemented in an all-optical setup~\cite{BLP_measure_exp_nat_phys_2011}.

In an experimental implementation of the BLP measure~(\ref{eq_blp_measure}), the trace distance $\mathrm{D}\left(\rho_1(t),\rho_2(t)\right)$ of a certain optimal initial state pair should be continuously monitored to depict its time profile and accurately identify local extrema. Consequently, a substantial amount of experimental raw data with sufficiently fine time resolution is required, particularly when the underlying dynamics are a priori unknown, as shown in Fig.~\ref{fig_nm_measure_illustration}\textbf{b}. Achieving this necessitates considerable experimental effort; otherwise, a coarse temporal resolution may prevent the precise identification of local extrema, leading to an unreliable estimation of the non-Markovianity, as shown in Fig.~\ref{fig_nm_measure_illustration}\textbf{c}. Here, we aim to significantly relax these stringent experimental requirements by leveraging the power of quantum machine learning. Unlike previous approaches~\cite{tancara_qsvm_non-mark_pra_2023, fanchini_svm_non-mark_pra_2021}, where the training features consist of the expectation values of the three Pauli matrices at a single fixed time, we explore how non-Markovianity can be estimated directly from the temporal correlations in sparse data, even in the absence of precise local extrema.

\subsection{Generation of training data}\label{sec: data generation}

In training supervised learning algorithms, a crucial stage is to generate training data with reliable labels. To demonstrate the feasibility of our approach, we consider the pure dephasing dynamics described by the total Hamiltonian
\begin{equation}
\widehat{H}_\mathrm{T}=\widehat{H}_\mathrm{S}+\widehat{H}_\mathrm{E}+\widehat{H}_\mathrm{I}
\label{eq_total_hamiltonian_s-b_model}
\end{equation}
with the system Hamiltonian $\widehat{H}_\mathrm{S}=\hbar\omega_0\hat{\sigma}_z/2$, the environment Hamiltonian
$\widehat{H}_\mathrm{E}=\sum_{\vec{k}}\hbar\omega_{\vec{k}}\hat{b}_{\vec{k}}^\dagger\hat{b}_{\vec{k}}$, and the interaction Hamiltonian
\begin{equation}
\widehat{H}_\mathrm{I}=\sum_{j=\uparrow,\downarrow}|j\rangle\langle j|\otimes\widehat{B}_j.
\label{eq_interaction_hamiltonian}
\end{equation}
Given the unitary time evolution operator $\widehat{U}^\mathrm{I}(t)=\mathcal{T}\left\{\exp\left[(-i/\hbar)\int_0^{t}\widehat{H}^\mathrm{I}_\mathrm{I}(\tau)d\tau\right]\right\}$ in the interaction picture, the qubit state at a later time $t$ is given by $\rho_\mathrm{S}(t)=\Tr_\mathrm{E}[\widehat{U}^\mathrm{I}(t)\rho_\mathrm{T}(0)\widehat{U}^{\mathrm{I}\dagger}(t)]$ for an initial total state $\rho_\mathrm{T}(0)$. In general, the environmental operator $\widehat{B}_j$ can be any Hermitian operator acting on the environmental Hilbert space. For example, let $\widehat{B}_\uparrow=-\widehat{B}_\downarrow=\sum_{\vec{k}}\hbar(g_{\vec{k}}\hat{b}_{\vec{k}}^\dagger+g_{\vec{k}}^\ast\hat{b}_{\vec{k}})$, which recovers the conventional spin-boson model with a real dephasing factor $\phi(t)=\bra{\uparrow}\rho_\mathrm{S}(t)\ket{\downarrow}/\bra{\uparrow}\rho_\mathrm{S}(0)\ket{\downarrow}$.

Here, we consider the biased spin-boson model~\cite{hongbin_cher_sr_2021}, a slight generalization of the conventional model, where the environmental operators are replaced by
\begin{equation}
\widehat{B}_j=\sum_{\vec{k}}\hbar(g_{j,\vec{k}}\hat{b}_{\vec{k}}^\dagger+g_{j,\vec{k}}^\ast\hat{b}_{\vec{k}}). \label{eq_b_s-b_model_env_operator}
\end{equation} Note that the coupling constants $g_{j,\vec{k}}$ depend on $j$ and vary accordingly with it. Assuming that the environment is in thermal equilibrium at temperature $T$, and the coupling constants are of equal magnitude, differing only by a finite relative phase, i.e.,
$g_{\downarrow,\vec{k}}=g_{\uparrow,\vec{k}}e^{i\varphi}$, then this model admits an analytical solution, where the dynamics of the qubit system is governed by a complex dephasing factor
\begin{equation}
\phi(t)=\exp\left[-i\vartheta(t)-\Phi(t)\right],
\label{eq_bsbm_dephasing_factor}
\end{equation}
where $\vartheta(t)=2\sin\varphi\int_0^\infty[\mathcal{J}(\omega)/\omega^2](1-\cos\omega t)d\omega$ and
$\Phi(t)=2(1-\cos\varphi)\int_0^\infty[\mathcal{J}(\omega)/\omega^2]\coth(\hbar\omega/2k_\mathrm{B}T)(1-\cos\omega t)d\omega$ incorporates the information about the interaction and the environmental density of states in terms of the spectral density $\mathcal{J}(\omega)=\sum_{\vec{k}}|g_{\vec{k}}|^2\delta(\omega-\omega_{\vec{k}})$. We specifically consider the family of super-Ohmic spectral densities
\begin{equation}
\mathcal{J}^{(s)}(\omega)=\eta\frac{\omega^s}{\omega_\mathrm{c}^{s-1}}e^{-\frac{\omega}{\omega_\mathrm{c}}}
\end{equation}
with $s>1$, $\eta$ describing the coupling strength, and $\omega_\mathrm{c}$ being the cut-off frequency. Detailed solutions and expressions of the qubit dynamics are presented in Methods.

For pure dephasing dynamics with dephasing factor $\phi(t)$, the trace distance of an optimal state pair equals to $\vert\phi(t)\vert$, i.e., $\mathrm{D}\left(\rho_1(t),\rho_2(t)\right)=\vert\phi(t)\vert$. Consequently, estimating the non-Markovianity $\mathcal{M}$ in Eq.~(\ref{eq_blp_measure}) of pure dephasing dynamics amounts to estimating the revival of the profile of $\vert\phi(t)\vert$. Since $\phi(t)$ in Eq.~(\ref{eq_bsbm_dephasing_factor}) has an analytical solution, the corresponding non-Markovianity $\mathcal{M}$ can be efficiently estimated. We uniformly sample Ohmicity $s\in[1.1,6.5]$ and temperature $T\in[0.5,4.5]$, generating 400 samples for training and 200 for testing. To realize the experimental sparse temporal data, we specify 10 time instances $\{t_1,\ldots,t_{10}\}$ of equal spacing and construct the feature $\vec{x}_{s,T}=\{\Re[\phi(t_1)],\Im[\phi(t_1)],\ldots,\Re[\phi(t_{10})],\Im[\phi(t_{10})]\}$ of length 20, which corresponds to the label $\mathcal{M}(s,T)$ for each sample $(s,T)$. The distribution of the generated non-Markoviantiy $\mathcal{M}$ is presented in Methods.

Implementing the numerical simulations of the aforementioned quantum models employing 20 qubits is computationally intensive. To enhance the efficiency of our numerical simulations and eliminate potential redundancy in high-dimensional feature space, the feature length is reduced to 10 using principal component analysis~\cite{PCA}. Additionally, to enhance the distinguishability of the data samples, as well as the trainability of the models, we also perform min-max normalization on a component-wise basis across the entire dataset to transform each component $x_{s,T}^{(\mu)}$ into the interval $[0,1]$.

\subsection{Numerical results}\label{sec:numerical results}

\begin{figure*}[ht]
    \centering \includegraphics[width=\textwidth]{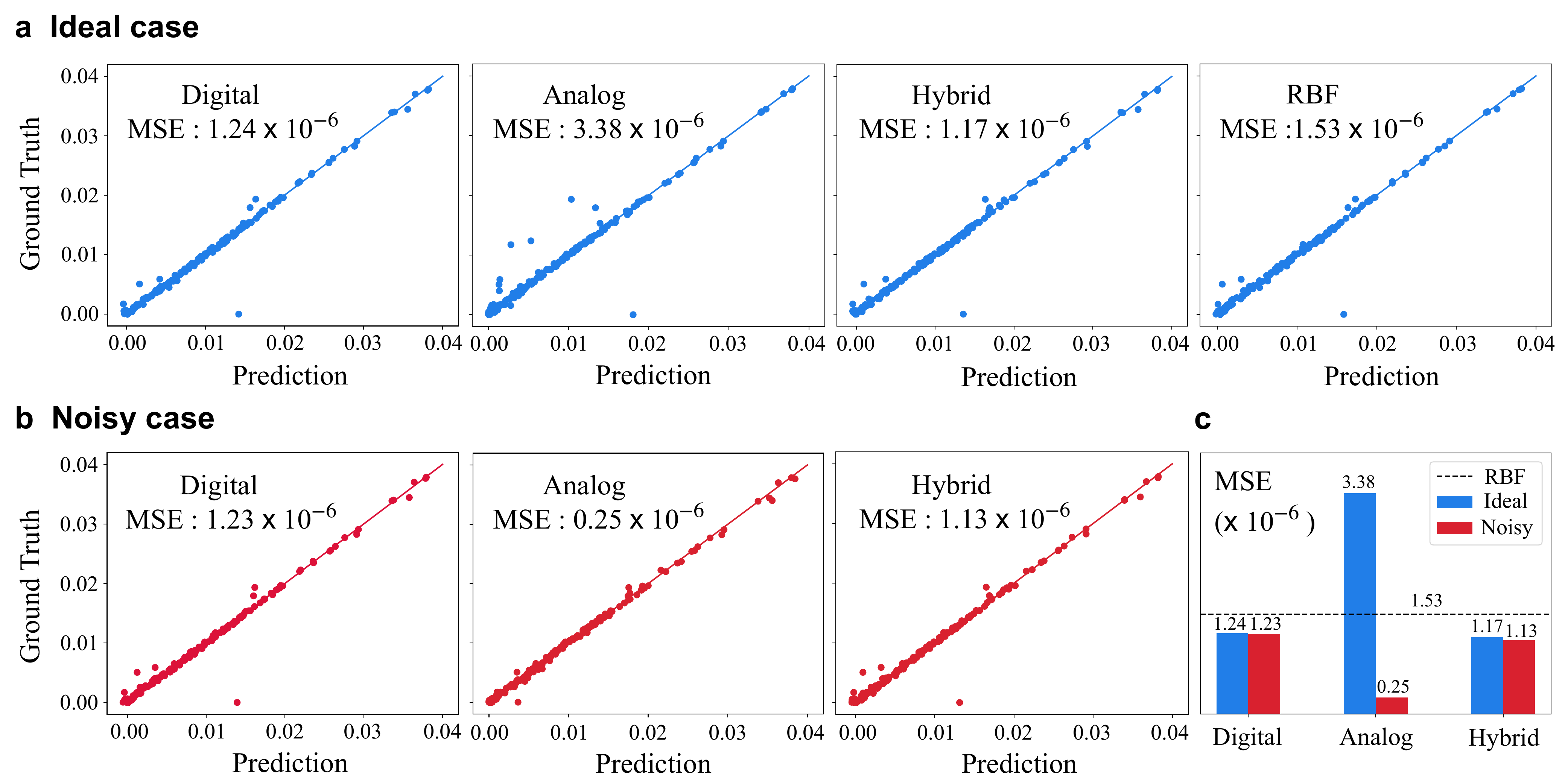}
    \caption{
    \textbf{Prediction of the non-Markovianity by quantum and RBF models.} We demonstrate the performance of the three quantum models, along with the RBF model, for the \textbf{a} ideal and \textbf{b} noisy cases in estimating the non-Markovianity from sparse temporal data. Most predictions align closely with the ground truth, demonstrating outstanding overall accuracy. Additionally, the noise-enhanced performance can also be observed in this case. \textbf{c} The MSE of the three quantum models is summarized with a histogram. In the presence of the noise, the performance of the digital quantum kernel remains unchanged, while the developed quantum kernel improves. This performance improvement is particularly prominent for the analog quantum model.}
    \label{fig:NM regression result}
\end{figure*}
\begin{figure*}[ht]
    \centering \includegraphics[width=\textwidth]{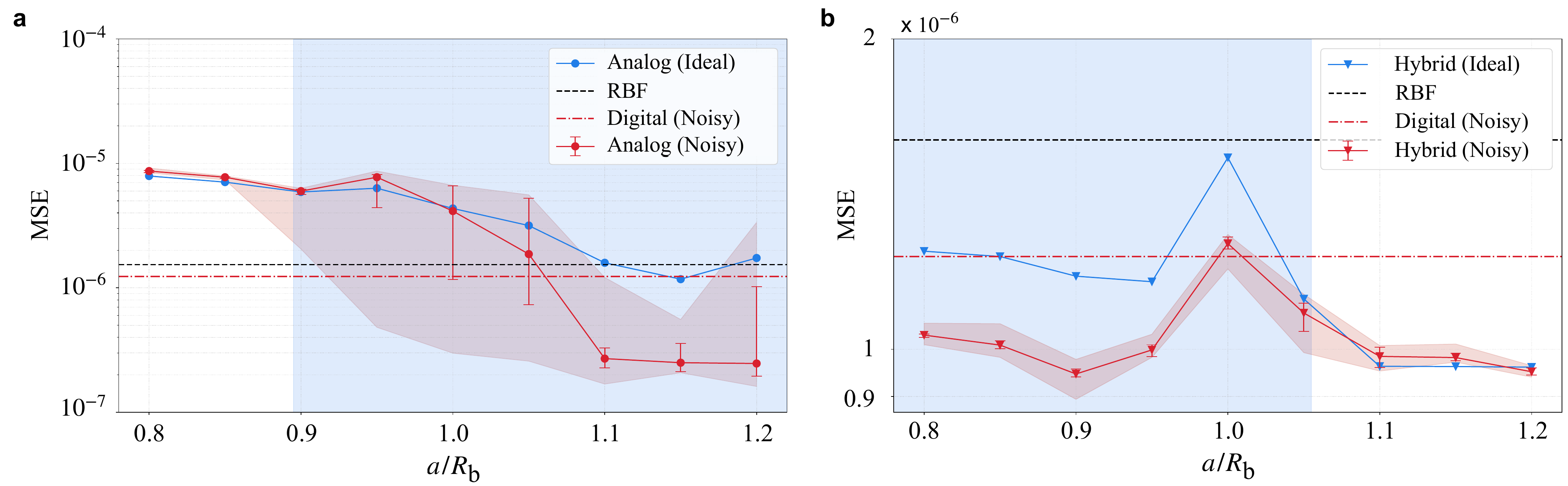}
    \caption{ 
    \textbf{Enhancement of performance in the presence of noise at various interatomic distances.} We show the performance of the \textbf{a} analog and \textbf{b} hybrid models in both ideal and noisy settings, along with the noisy digital and RBF models for comparison. Error bars are estimated from 11 independent simulations, with the center markers indicating the median and the bars representing the first and third quartiles. The red-shaded area represents the full range of the data, while the blue-shaded region highlights the regime where performance enhancement is observed. The performance of both analog and hybrid models can be enhanced in the presence of noise at interatomic distances of $a\geq 0.9~R_\mathrm{b}$ and $a\leq 1.05~R_\mathrm{b}$, respectively. For an appropriate choice of interatomic distance, the analog and hybrid quantum models achieve performance comparable to or even superior to that of the noisy digital and RBF models; meanwhile, the magnitude of the performance enhancement of the analog quantum model is more pronounced than that of the hybrid one.}
    \label{fig: NM ideal vs noise}
\end{figure*}
We utilize QuantumToolbox.jl~\cite{Alberto2025} to implement numerical simulations of the quantum kernel functions in both ideal and noisy settings. Given our computational resources, we set $M=1{,}000$ for each noise ensemble during our simulations. The corresponding quantum models, along with a classical RBF model for comparison, are then trained using Scikit-learn~\cite{sklearn} via a grid search over hyperparameters $C$ and $\epsilon$.

The numerical results are shown in Fig.~\ref{fig:NM regression result}, where the performance is evaluated by mean squared error (MSE), with lower values indicating higher accuracy. At an interatomic distance of $a=1.05$~$R_\mathrm{b}$, all models achieve low MSE values, and most predictions align closely with the ground truth, indicating robust overall performance. In particular, as shown in Fig.~\ref{fig:NM regression result}\textbf{a}, the digital and hybrid quantum kernels outperform the RBF model. This suggests that these quantum machine learning models are competitive with the classical RBF kernel on this task, offering a viable approach to substantially relax the stringent experimental demands.

We further investigate the effects of the noise on the performance of the three quantum models, with the numerical results shown in Fig.~\ref{fig:NM regression result}\textbf{b}. Typically, noise in quantum devices is considered detrimental to performance. However, our numerical results reveal a counterintuitive trend: the presence of noise in the quantum feature map can improve the performance of the quantum models on this task. This performance enhancement, as summarized in Fig.~\ref{fig:NM regression result}\textbf{c}, is more pronounced for the analog quantum model. To further reveal the ubiquity of this phenomenon, we extensively train a range of analog and hybrid quantum models at various interatomic distances $a$, performing 11 independent simulations for each noisy model at each distance. The numerical results are presented in Fig.~\ref{fig: NM ideal vs noise}, where we report the median, first quartile, third quartile, and the full range of the data. In the presence of noise, the performance enhancement of the analog model (Fig.~\ref{fig: NM ideal vs noise}\textbf{a}) can be observed when the interatomic distance exceeds $0.9~R_\text{b}$, and the median of the MSE is lower than that of the ideal case for $a\geq1.05$~$R_\mathrm{b}$. In particular, each sample of the noisy analog model shows pronounced performance enhancement when the interatomic distance is set to $1.1~R_\text{b}$ or $1.15~R_\text{b}$. By contrast, the magnitude of the enhancement of the noisy hybrid model (Fig.~\ref{fig: NM ideal vs noise}\textbf{b}) is smaller, but is observed consistently when the interatomic distance is set below $1.05~R_\text{b}$. Moreover, in both ideal and noisy settings, each model can achieve performance superior to that of the noisy digital and RBF models when the interatomic distance is appropriately chosen, demonstrating that the developed quantum kernels remain competitive with the classical kernel even in the presence of noise.

\section{Noise-enhanced performance, expressivity, model complexity, and benign overfitting}

\begin{figure*}
    \centering
    \includegraphics[width=\textwidth]{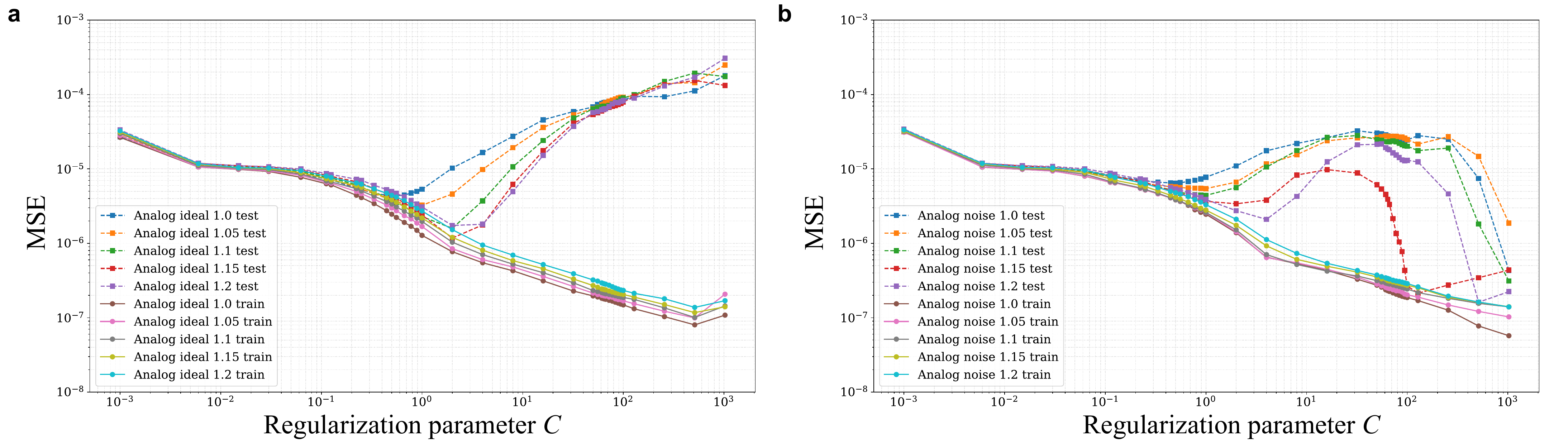}
    \caption{
    \textbf{Regularization paths of the analog quantum model.} We compare the regularization paths of the analog quantum model in the \textbf{a} ideal and \textbf{b} noisy settings when performance enhancement occurs. The training and test MSEs are represented by the solid and dashed curves, respectively. In the high-complexity regime, the analog quantum model in both settings closely fits the training data. However, the gap between the training and test MSEs widens for the ideal setting, while the noisy model can keep the test MSE under control.}
    \label{fig: analog regularization path}
\end{figure*}
\begin{figure*}
    \centering
    \includegraphics[width=\textwidth]{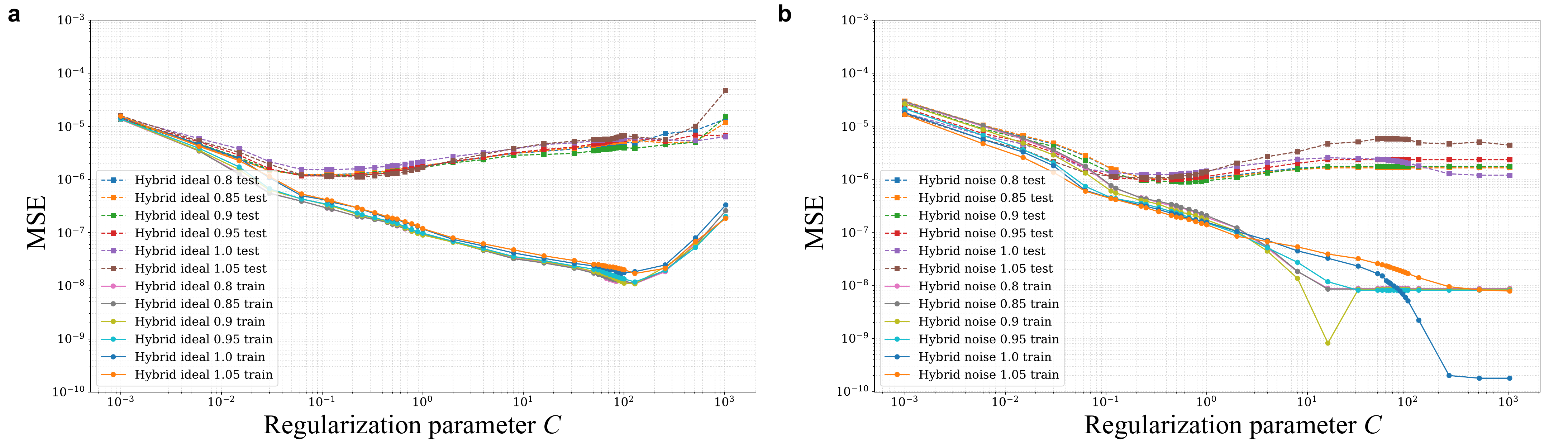}
    \caption{
    \textbf{Regularization paths of the hybrid quantum model.} We compare the regularization paths of the hybrid quantum model in the \textbf{a} ideal and \textbf{b} noisy settings when performance enhancement occurs. The training and test MSEs are represented by the solid and dashed curves, respectively. Despite closely fitting the training data, the ideal hybrid quantum model keeps test MSE under control, and the presence of noise can further preserve this feature in the high-complexity regime, resulting in a lower test MSE.}
    \label{fig: hybrid regularization path}
\end{figure*}
\begin{figure*}
    \centering
    \includegraphics[width=\textwidth]{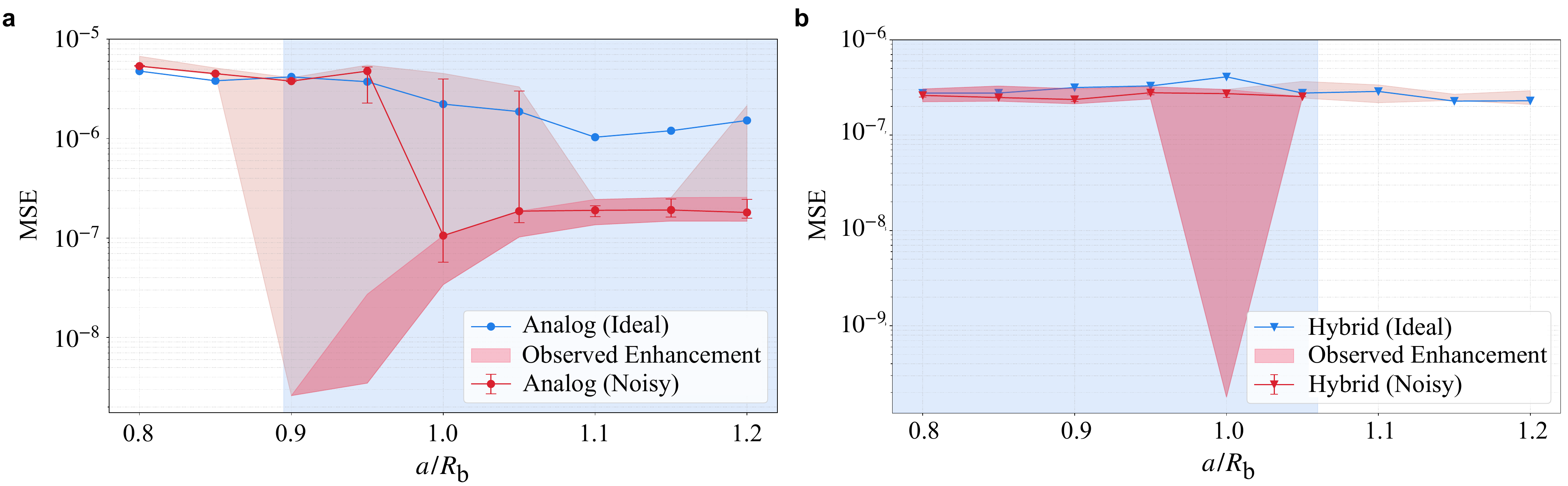}
    \caption{
    \textbf{Training MSE in the presence of noise at various interatomic distances.} We show the training MSE of the \textbf{a} analog and \textbf{b} hybrid quantum models in both noisy and ideal settings. Error bars are estimated from 11 independent simulations, with the center markers indicating the median and the bars representing the first and third quartiles. The red-shaded region represents the full range of the data, and the darker red-shaded region highlights the regime where performance enhancement occurs. Within this enhancement regime, the training MSE of the noisy quantum models is consistently lower than, or comparable to, that of their ideal counterparts.}
    \label{fig: ideal vs noise train MSE}
\end{figure*}
\begin{figure*}[ht]
    \centering \includegraphics[width=\textwidth]{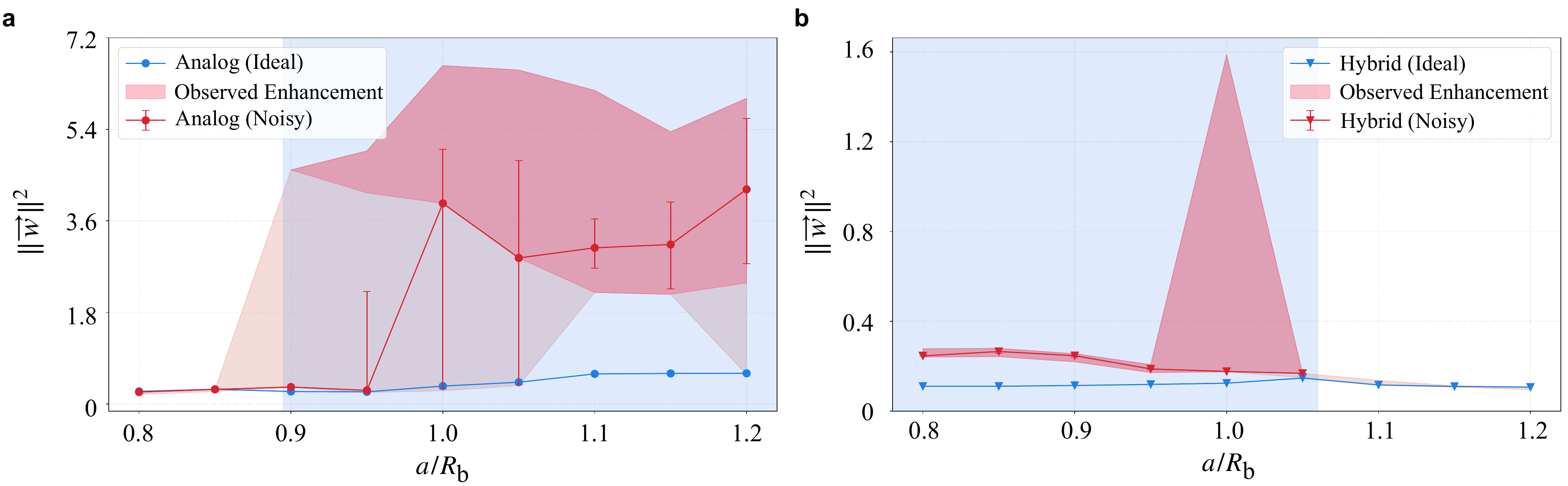}
    \caption{
    \textbf{Enhancement of the squared norm of weight $\Vert\vec{w}\Vert^2$ in the presence of noise at various interatomic distances.} We show the model complexity quantified by the squared norm of weight $\Vert \vec{w}\Vert^2$ of the decision function $f(\vec{x})$ for the \textbf{a} analog and \textbf{b} hybrid quantum models. Error bars are estimated from 11 independent simulations, with the center markers indicating the median and the bars representing the first and third quartiles. The red-shaded region represents the full range of the data, and the darker red-shaded region highlights the regime where performance enhancement occurs. In this enhancement regime, $\Vert\vec{w}\Vert^2$ is consistently larger than the model complexity observed in their ideal counterparts. Consequently, these noisy quantum models can support higher model complexity while keeping the test MSE under control.
    }
    \label{fig: decision func complexity}
\end{figure*}
\begin{figure*}
    \centering
    \includegraphics[width=\textwidth]{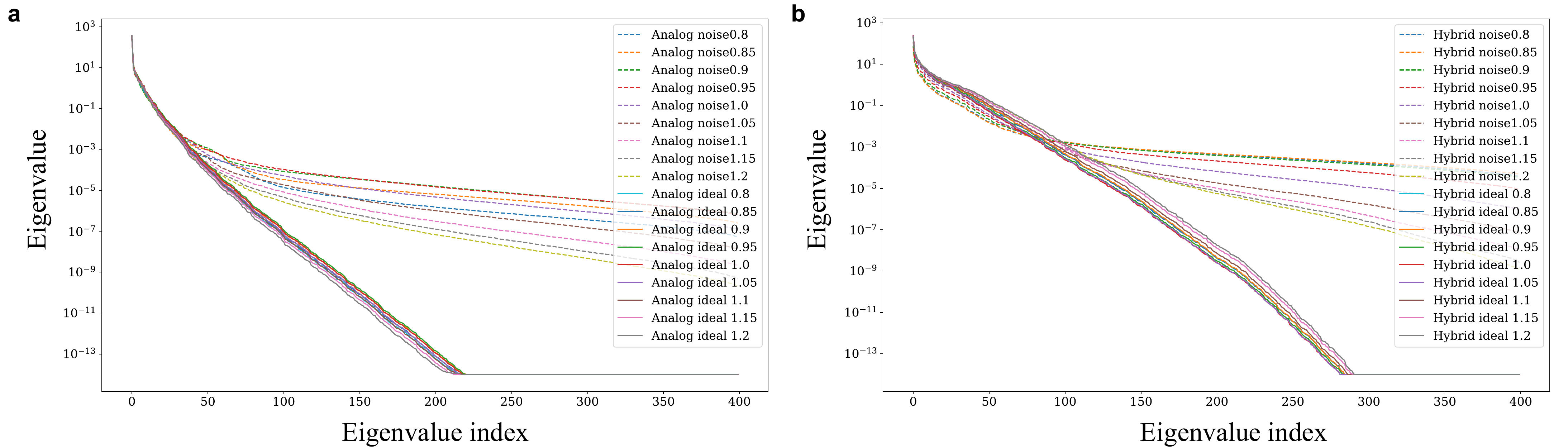}
    \caption{
    \textbf{Eigenspectrum analysis of the ideal and noisy kernel matrices.} We analyze the eigenspectra of the \textbf{a} analog and \textbf{b} hybrid quantum kernel matrices in both ideal (solid curve) and noisy (dashed curve) settings, with the eigenvalues sorted in descending order. In the ideal setting, the eigenspectra of both quantum kernel matrices exhibit an exponential decay. By contrast, the presence of noise lifts the near-zero eigenvalues, so that the subdominant eigenvectors, which are effectively suppressed in the ideal setting, now contribute appreciably. 
    }
    \label{fig: kernel spectrum}
\end{figure*}

We further explore the mechanisms underlying the observed performance enhancement. We observe that the presence of noise can help the analog and hybrid quantum models keep the test MSE under control despite closely fitting the training data. To demonstrate this, we track the regularization paths of both quantum models, showing their training and test performance as model complexity increases. The results are shown in Figs.~\ref{fig: analog regularization path} and \ref{fig: hybrid regularization path} for analog and hybrid quantum models, respectively. In the ideal setting, both quantum models overfit the training data as model complexity increases, causing the gap between the training and test MSEs to widen in the high-complexity regime. However, when the noisy quantum models outperform their ideal counterparts, they admit a higher-complexity solution that keeps the test MSE under control despite closely fitting the training data. Consequently, these noisy quantum models exhibit higher model complexity and a training MSE lower than or comparable to that of their ideal counterparts. To verify this behavior, we examine the training MSE and the complexity of the decision function for both quantum models in Figs.~\ref{fig: ideal vs noise train MSE} and \ref{fig: decision func complexity}, where the complexity is quantified by the squared Euclidean norm of the weights, $\Vert\vec{w}\Vert^2$. These characteristics are consistently observed in noisy quantum models whenever the performance enhancement occurs, as indicated in the darker red-shaded areas in Figs.~\ref{fig: ideal vs noise train MSE} and \ref{fig: decision func complexity}.

This improved generalization is accompanied by a distinct structural change in the kernel eigenspectrum, as shown in Fig.~\ref{fig: kernel spectrum}. In the ideal setting, the eigenspectra of the analog and hybrid quantum kernel matrices decay exponentially. In the presence of noise, the tails of both eigenspectra are lifted, so that the subdominant eigenvectors, which are effectively suppressed in the ideal setting, now contribute appreciably. The change of the kernel eigenspectrum behavior, combined with a close fit to the training data and a well-controlled test MSE, closely resembles the tempered overfitting phenomenon characterized for kernel ridge regression~\cite{Mallinar2022}, in which the tail behavior of the kernel eigenspectrum is associated with whether the model overfits catastrophically or in a tempered manner. However, our utilized algorithm, SVR, differs from kernel ridge regression (KRR) because the SVR uses an $\varepsilon$-insensitive loss and generally does not interpolate the data. Therefore, we invoke this correspondence descriptively, suggesting that the noise induces a mechanism analogous to tempered overfitting in the regime where the performance enhancement is observed.

\section{Conclusion and discussion}

Due to their feasibility, conventional digital quantum kernels have been extensively applied across a wide variety of fields and experimentally realized on various quantum computing platforms. However, hindered by significant noise in contemporary quantum devices, it remains challenging for these models to outperform their classical counterparts in solving practical problems. We propose to circumvent this limitation by constructing quantum kernels in the context of analog quantum computing. We construct two possible architectures. The analog quantum kernel encodes the features into the local energy difference of a Rydberg atom system. Meanwhile, the hybrid quantum kernel adopts a digital-analog ansatz to encode the features via single-qubit gates and replaces the entangling layer with Hamiltonian evolution.

To evaluate the practical utility of these quantum kernels, we apply them to the physically motivated application of estimating the non-Markovianity from sparse temporal data. Most measures of non-Markovianity hinge on the technique of quantum process tomography with sufficiently fine time resolution, requiring extensive experimental resources, particularly when dealing with complex dynamics. We show that, by leveraging the power of quantum machine learning models, stringent experimental requirements can be substantially alleviated, and the non-Markovianity can be estimated with high accuracy from sparse data. 

Moreover, we investigate the behavior of these quantum models in the presence of operational noise and depolarizing noise. Although noise is usually regarded as detrimental to performance, our numerical results reveal that noisy quantum models can counterintuitively outperform their ideal counterparts when the interatomic distances are set to $a \geq 0.9~R_\mathrm{b}$ and $a \leq 1.05~R_\mathrm{b}$ for analog and hybrid quantum models, respectively. This demonstrates that these quantum models can remain highly competitive, and even thrive, in the presence of noise.

To explore the underlying mechanism of this phenomenon, we perform a quantitative analysis of these quantum models, revealing that the noise can help the analog and hybrid quantum models keep the test MSE under control despite closely fitting the training data. Consequently, these noisy quantum models exhibit higher model complexity and training MSEs lower than or comparable to those of their ideal counterparts. Additionally, we find that the noise lifts the near-zero tail of the kernel eigenspectrum, replacing the characteristic exponential decay of the ideal setting with a slower one. Since this behavior resembles the tempered overfitting in the KRR~\cite{Mallinar2022}, we suggest that the noise induces behavior analogous to tempered overfitting in the regime where the enhancement is observed. However, since SVR utilizes an $\varepsilon$-insensitive loss and does not interpolate the data in general, we invoke this correspondence only descriptively and leave the rigorous theoretical proof of this conjecture for future work.

Our work naturally opens several promising directions for future research. We have shown that the experimental efforts for estimating the non-Markovianity of complex dynamics can be substantially alleviated using quantum machine learning models. It would be interesting to investigate whether such models can also be beneficial for reducing experimental effort in other experiments that typically require intensive measurements. Regarding the noise-enhancement phenomenon, we note that several recent studies have also reported instances where noise may be counterintuitively beneficial to quantum machine learning models~\cite{Linghua2026, Domingo2023, Liu2025, Du2021}. However, other studies have demonstrated that noise remains harmful to quantum machine learning models in different contexts. For instance, in variational quantum algorithms, noise can induce a limit set for the loss function and reduce the quantum state to a fixed point of the noise model~\cite{Singkanipa2025, Stilck2021}. This tension highlights the need for a comprehensive investigation into the role of noise across different kinds of quantum machine learning algorithms. Ultimately, identifying the precise conditions under which noise can be systematically exploited as a beneficial resource for a given quantum feature map will substantially deepen our understanding of quantum machine learning models.
\section*{Data availability}

The data that support the findings of this study are available upon reasonable request from the corresponding authors.
\section*{Acknowledgments}

This work is supported by the National Science and Technology Council, Taiwan, with Grants No. MOST 111-2112-M-006-015-MY3, NSTC 114-2112-M-006-015-MY3, and NSTC 114-2628-M-006-002-MY4, partially by the Higher Education Sprout Project, Ministry of Education to the Headquarters of University Advancement at NCKU, and partially by the National Center for Theoretical Sciences, Taiwan.

\onecolumngrid
\appendix

\section{The effect of the depolarizing noise on the quantum kernel}\label{appendix A}
Here, we analyze the effect of depolarizing noise on the quantum kernel. Since a fidelity quantum kernel estimates similarity through the Hilbert-Schmidt inner product in the Hilbert space, applying the depolarizing noise to the quantum state is equivalent to shifting and rescaling the quantum kernel function as a function of the depolarizing factor $p$. The resulting noisy quantum kernel function reads

\begin{align}
     \tilde{\kappa}_\mathrm{Q}^\mathrm{dp}(\vec{x}_k,\vec{x}_j) &= \Tr\left[[(1-p)\rho(\data{j}) +\frac{p}{2^d}\widehat{I}][(1-p)\rho(\data{k}) +\frac{p}{2^d}\widehat{I}]\right] \\
    &= (1-p)^2\Tr[\rho(\data{j})\rho(\data{k})] + 2\frac{p(1-p)}{2^d} + \frac{p^2}{2^d} \\
    &= (1-p)^2\kappa_\mathrm{Q}(\vec{x}_j,\vec{x}_k) + \frac{p(2-p)}{2^d},
\end{align}

where $2^d$ is the dimension of the Hilbert space. Since the shift is data-independent, the kernel matrix under the depolarizing noise can be written as 
\begin{equation}
    \widetilde{K}_\mathrm{Q}^\mathrm{dp} = (1-p)^2\widehat{K}_\mathrm{Q} + \frac{p(2-p)}{2^d}\widetilde{J},
\end{equation}
where $\widetilde{J}=\vec{1}\vec{1}^\mathrm{T}$ is the matrix of ones and $\vec{1}$ is the vector of ones. Applying this noisy kernel matrix to the SVR algorithm in Eq.~\eqref{eq_optimization_dual}, the loss function only differs by a rescaling of the hyperparameter $C$ in the first term. Specifically, the first term of the loss under the depolarizing noise reads 
\begin{align}
        L^{\mathrm{dp}}_1 &= (\vec{\alpha}-\vec{\alpha}^*)^\mathrm{T}\widetilde{K}_\mathrm{Q}^\mathrm{dp} (\vec{\alpha}-\vec{\alpha}^*)\\
        &=(\vec{\alpha}-\vec{\alpha}^*)^\mathrm{T}(c_0\widehat{K}_\mathrm{Q} + c_1\vec{1}\vec{1}^\mathrm{T} )(\vec{\alpha}-\vec{\alpha}^*) \\
        &= (\vec{\alpha}-\vec{\alpha}^*)^\mathrm{T}(c_0\widehat{K}_\mathrm{Q})(\vec{\alpha}-\vec{\alpha}^*) + c_1(\vec{\alpha}-\vec{\alpha}^*)^\mathrm{T}\vec{1}\vec{1}^\mathrm{T}(\vec{\alpha}-\vec{\alpha}^*)\\
        &= c_0(\vec{\alpha}-\vec{\alpha}^*)^\mathrm{T}(\widehat{K}_\mathrm{Q})(\vec{\alpha}-\vec{\alpha}^*),
\end{align}
where we substitute $c_0=(1-p)^2$ and $c_1=p(2-p)/2^d$, and the third equality follows from the constraint $(\vec{\alpha}-\vec{\alpha}^*)^\mathrm{T}\vec{1} = 0$ in Eq.~\eqref{eq_optimization_dual}. Hence, the depolarizing noise leaves the training process unchanged, and the original loss can be recovered by rescaling the hyperparameter $C$. Moreover, when predicting unseen data, the decision function is also modified by a scaling factor, which is
\begin{align}
    f^{\mathrm{dp}} (\vec{x}) &= \sum_{j=1}^{N} (\alpha_j - \alpha_j^*)(c_0\kappa_\mathrm{Q}(\vec{x},\vec{x}_j)+ c_1)+b  \\
    &= c_0\sum_{j=1}^{N} (\alpha_j - \alpha_j^*)\kappa_\mathrm{Q}(\vec{x},\vec{x}_j)+ c_1\sum_{j=1}^{N} (\alpha_j - \alpha_j^*) + b \\
    &= c_0\sum_{j=1}^{N} (\alpha_j - \alpha_j^*)\kappa_\mathrm{Q}(\vec{x},\vec{x}_j) +b,
\end{align}
where the second equality again follows from the constraint in Eq.~\eqref{eq_optimization_dual}, and the overall scaling $c_0$ can be absorbed into the variables $\vec{\alpha}$ and $\vec{\alpha}^*$. This shows that the quantum SVR is robust to the depolarizing noise, since its effect can be compensated by searching over a larger grid for $ C$. We note that similar findings have been reported numerically \cite{icaart26}.

\section{Biased spin-boson model}\label{sec_bia_sbm}

Here we present the detailed calculations for the biased spin-boson models, as described by the Hamiltonian operators in Eqs.~(\ref{eq_total_hamiltonian_s-b_model})-(\ref{eq_b_s-b_model_env_operator}). In the interaction picture with respect to $\widehat{H}_\mathrm{S}+\widehat{H}_\mathrm{E}$, the interaction Hamiltonian in Eq.~(\ref{eq_interaction_hamiltonian}) is written as
\begin{equation}
\widehat{H}^\mathrm{I}_\mathrm{I}(t)=\sum_{j=\uparrow,\downarrow}|j\rangle\langle j|\otimes\widehat{B}_j^{\mathrm{I}}(t),
\end{equation}
where
\begin{equation}
\widehat{B}_j^{\mathrm{I}}(t)=\sum_{\vec{k}}\hbar(g_{j,\vec{k}}\hat{b}_{\vec{k}}^\dagger e^{i\omega_{\vec{k}} t}
+g_{j,\vec{k}}^\ast\hat{b}_{\vec{k}}e^{-i\omega_{\vec{k}} t}).
\end{equation}
Then the qubit and the bosonic environment will evolve jointly according to the unitary time evolution operator $\widehat{U}^\mathrm{I}(t)=\mathcal{T}\left\{\exp\left[(-i/\hbar)\int_0^{t}\widehat{H}^\mathrm{I}_\mathrm{I}(\tau)d\tau\right]\right\}$ with $\mathcal{T}$ being the time-ordering operator. The detailed calculation of $\widehat{U}^\mathrm{I}(t)$ can be found in Ref.~\cite{hongbin_cher_sr_2021}. Following we present the final results:
\begin{equation}
\widehat{U}^\mathrm{I}(t)=\sum_{j=\uparrow,\downarrow}|j\rangle\langle j|\otimes\prod_{\vec{k}}\exp\left[i|g_{j,\vec{k}}|^2\frac{\omega_{\vec{k}}t-\sin\omega_{\vec{k}}t}{\omega_{\vec{k}}^2}\right]
\widehat{D}[g_{j,\vec{k}}\beta_{\vec{k}}(t)],
\end{equation}
where $\widehat{D}[g_{j,\vec{k}}\beta_{\vec{k}}(t)]=\exp[g_{j,\vec{k}}\beta_{\vec{k}}(t)\hat{b}_{\vec{k}}^\dagger-g_{j,\vec{k}}^\ast\beta_{\vec{k}}^\ast(t)\hat{b}_{\vec{k}}]$ is the displacement operator and $\beta_{\vec{k}}(t)=(1-e^{i\omega_{\vec{k}}t})/\omega_{\vec{k}}$.

Assuming that there is no initial correlation between the system and the environment, i.e., $\rho_\mathrm{T}(0)=\rho_\mathrm{S}(0)\otimes\rho_\mathrm{E}(0)$, and $\rho_\mathrm{E}(0)=\exp[-\widehat{H}_\mathrm{E}/k_\mathrm{B}T]/Z$ is an equilibrium state at temperature $T$, the qubit state at a later time $t$ is given by $\rho_\mathrm{S}(t)=\Tr_\mathrm{E}[\widehat{U}^\mathrm{I}(t)\rho_\mathrm{T}(0)\widehat{U}^{\mathrm{I}\dagger}(t)]$. Due to the block diagonal form of the interaction Hamiltonian~(\ref{eq_interaction_hamiltonian}), the qubit undergoes pure dephasing dynamics characterized by a dephasing factor
\begin{eqnarray}
\phi(t)&=&\bra{\uparrow}\rho_\mathrm{S}(t)\ket{\downarrow}/\bra{\uparrow}\rho_\mathrm{S}(0)\ket{\downarrow}  \nonumber\\
&=&\Tr\left\{
\prod_{\vec{k}}\exp\left[-i|g_{\downarrow,\vec{k}}|^2\frac{\omega_{\vec{k}}t-\sin\omega_{\vec{k}}t}{\omega_{\vec{k}}^2}\right]
\widehat{D}[-g_{\downarrow,\vec{k}}\beta_{\vec{k}}(t)]
\prod_{\vec{k}}\exp\left[i|g_{\uparrow,\vec{k}}|^2\frac{\omega_{\vec{k}}t-\sin\omega_{\vec{k}}t}{\omega_{\vec{k}}^2}\right]\widehat{D}[g_{\uparrow,\vec{k}}\beta_{\vec{k}}(t)]\rho_\mathrm{E}(0)
\right\}.
\end{eqnarray}
Along with the two prescriptions $\widehat{D}[\beta_1]\widehat{D}[\beta_2]=\exp[(\beta_1\beta^\ast_2-\beta^\ast_1\beta_2)/2]\widehat{D}[\beta_1+\beta_2]$ and $\Tr\left\{\widehat{D}[\beta]\rho_\mathrm{E}(0)\right\}=\exp[-\coth(\hbar\omega/2k_\mathrm{B}T)|\beta|^2/2]$, the dephasing factor can be expressed analytically as
\begin{equation}
\phi(t)=\exp\left[-i\vartheta(t)-\Phi(t)\right],
\end{equation}
where
\begin{equation}
\vartheta(t)=2\sin\varphi\int_0^\infty[\mathcal{J}(\omega)/\omega^2](1-\cos\omega t)d\omega
\end{equation}
and
\begin{equation}
\Phi(t)=2(1-\cos\varphi)\int_0^\infty[\mathcal{J}(\omega)/\omega^2]\coth(\hbar\omega/2k_\mathrm{B}T)(1-\cos\omega t)d\omega.
\end{equation}
Note that, in the above expressions, for simplicity, we have assumed that $g_{\downarrow,\vec{k}}=g_{\uparrow,\vec{k}}e^{i\varphi}$.

If we further consider the family of super-Ohmic spectral densities
\begin{equation}
\mathcal{J}^{(s)}(\omega)=\eta\frac{\omega^s}{\omega_\mathrm{c}^{s-1}}e^{-\frac{\omega}{\omega_\mathrm{c}}}
\end{equation}
with $s>1$, then the above two integrals can be solved analytically as
\begin{equation}
\vartheta(t)=\eta\sin\varphi\Gamma(s-1)\left[2-\frac{\left(1+i\omega_\mathrm{c}t\right)^{s-1}+\left(1-i\omega_\mathrm{c}t\right)^{s-1}}{\left(1+\omega_\mathrm{c}^2t^2\right)^{s-1}}\right]
\end{equation}
and
\begin{eqnarray}
\Phi(t)&=&-\eta(1-\cos\varphi)\Gamma(s-1)\left[2-\frac{\left(1+i\omega_\mathrm{c}t\right)^{s-1}+\left(1-i\omega_\mathrm{c}t\right)^{s-1}}{\left(1+\omega_\mathrm{c}^2t^2\right)^{s-1}}\right] \nonumber\\
&&+2(1-\cos\varphi)\eta\Gamma(s-1)\left(\frac{k_\mathrm{B}T}{\hbar\omega_\mathrm{c}}\right)^{s-1}\left[2\zeta\left(s-1,\frac{k_\mathrm{B}T}{\hbar\omega_\mathrm{c}}\right)
-\zeta\left(s-1,\frac{k_\mathrm{B}T}{\hbar\omega_\mathrm{c}}(1+i\omega_\mathrm{c}t)\right)-\zeta\left(s-1,\frac{k_\mathrm{B}T}{\hbar\omega_\mathrm{c}}(1-i\omega_\mathrm{c}t)\right)\right], \nonumber\\
\end{eqnarray}
where $\zeta(s,q)=\sum_{n=0}^\infty(q+n)^{-s}$ is the Hurwitz zeta function. In the construction of the dataset, the cut-off frequency $\omega_\mathrm{c}$, the coupling strength $\eta$, the Boltzmann constant $k_\mathrm{B}$, and the reduced Planck constant $\hbar$ are all set to unity.

\section{The distribution of the non-Markovianity dataset}
\begin{figure*}
    \centering
    \includegraphics[width=\textwidth]{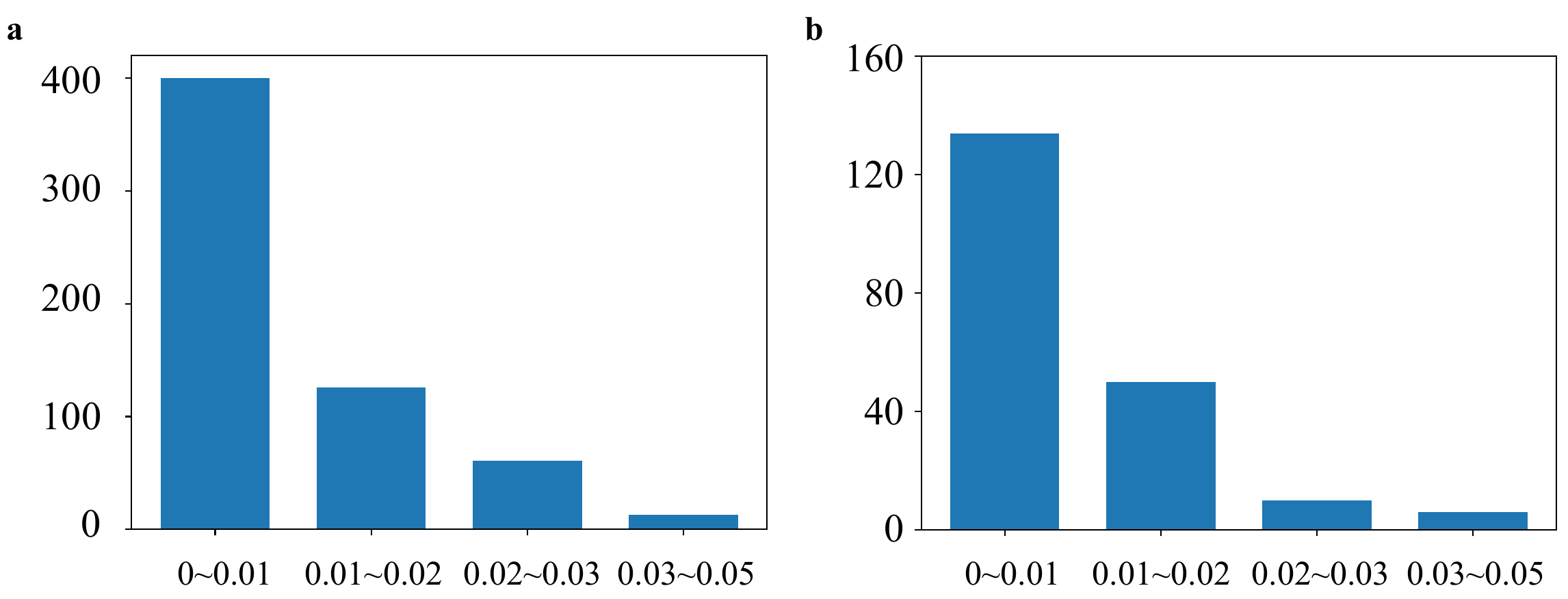}
    \caption{\textbf{Label distribution of the training and testing dataset.} We show the distribution of the label in the \textbf{a} training and \textbf{b} testing dataset. In both datasets, most of the labels lie in $[0,0.01]$, and the rest of the labels are distributed into the intervals $[0 .01,0.02]$, $[0.02,0.03]$, and $[0.03,0.05]$.  }
    \label{fig: dataset distribution}
\end{figure*}

The distribution of the training and testing dataset is shown in Fig.~\ref{fig: dataset distribution}. The label values in our dataset lie in the interval $[0,0.05]$. In both datasets, most of the labels lie in $[0,0.01]$, and the rest of the labels are distributed into the intervals $[0 .01,0.02]$, $[0.02,0.03]$, and $[0.03,0.05]$.

\section{Additional analyses on the noisy analog and hybrid quantum models}
\begin{figure*}
    \centering
    \includegraphics[width=\textwidth]{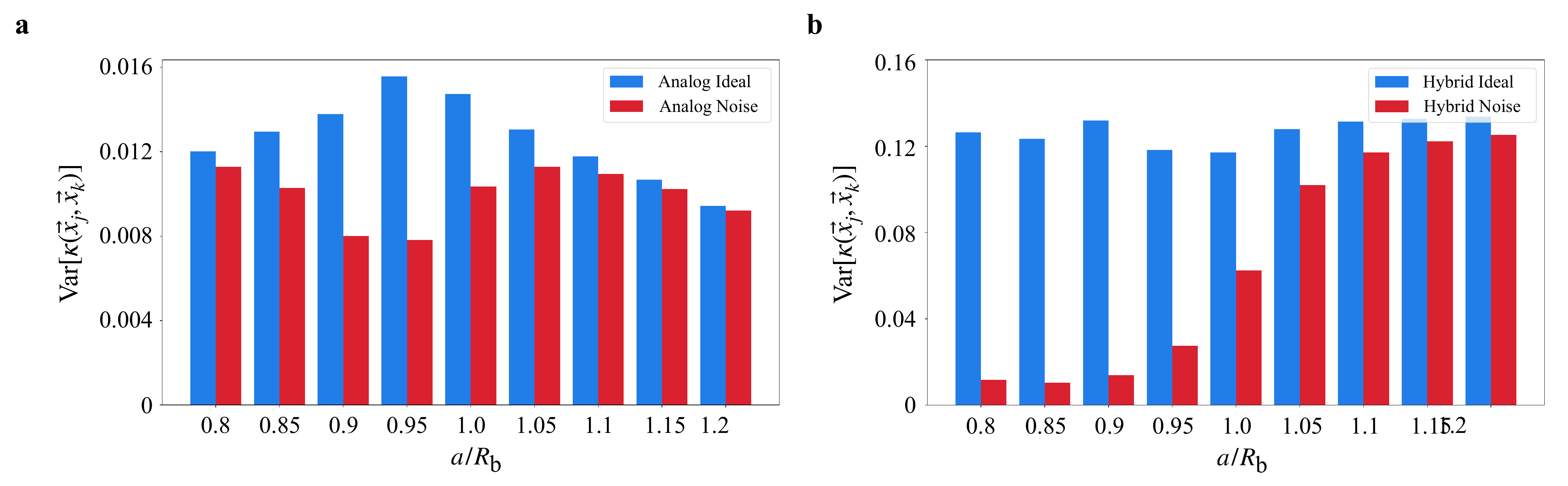}
    \caption{
    \textbf{Variance of the kernel elements in the analog and hybrid quantum kernel matrices.} We compare the $\mathrm{Var}[\tilde{\kappa}(\vec{x}_k,\vec{x}_j)]$ of the analog and hybrid quantum model in both ideal and noisy settings. The lower variance of the analog quantum kernel indicates that its values are more concentrated around the mean than those of the hybrid quantum kernel. Therefore, the experimental cost of the analog quantum kernel is higher than that of the hybrid quantum kernel in terms of the shot number. In the presence of noise, both quantum kernels exhibit reduced kernel variance, indicating increased experimental cost. In particular, the magnitude of the variance reduction is more pronounced in the hybrid quantum kernel, showing that its variance is more susceptible to the noise.
    }
    \label{fig: kernel variance}
\end{figure*}

\begin{figure*}
    \centering
    \includegraphics[width=\textwidth]{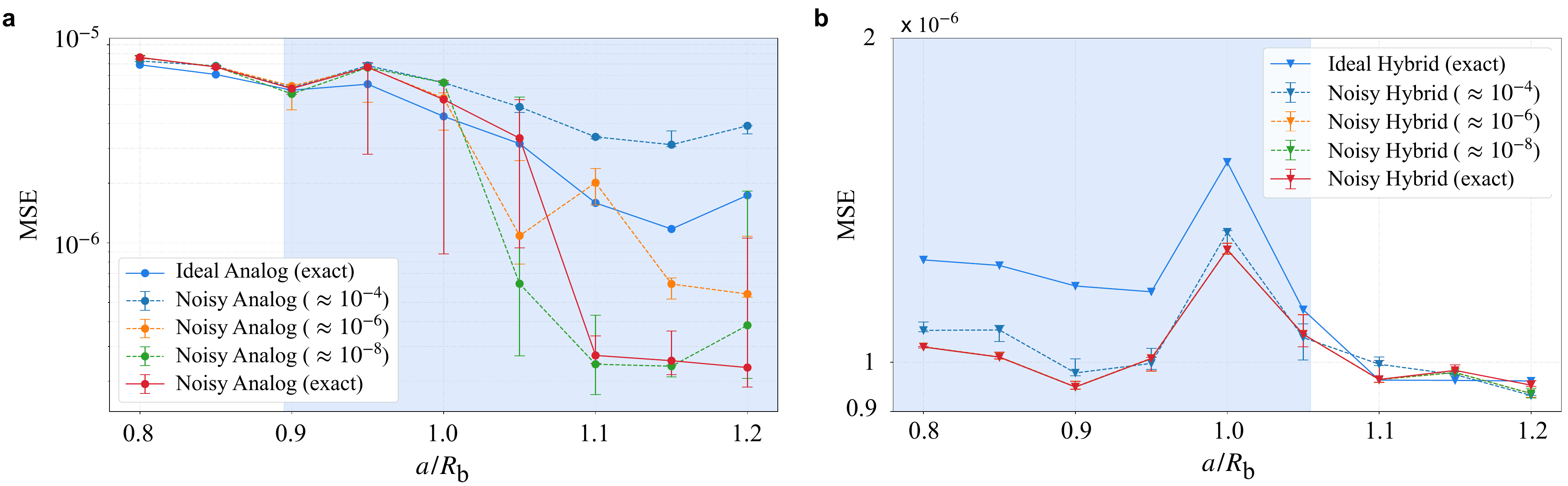}
    \caption{
    \textbf{Test performance under the shot noise of the analog and hybrid quantum kernel models.} We investigate the performance of the analog and hybrid quantum model under the effect of the shot noise. The points with a dashed curve show the performance of the noisy analog and hybrid quantum models under the shot noise, where the elements of the associated kernel matrix are rounded to different decimal places. The points with solid curves indicate the performance of each quantum model in the absence of shot noise. As predicted by Fig.~\ref{fig: kernel variance}, the analog quantum model requires a high precision ($\approx 10^{-8}$) to reproduce the performance under exact estimation, whereas the hybrid quantum model has already approached the performance under exact examination with precision $\approx 10^{-4}$.  
    }
    \label{fig: performance under shotnoise}
\end{figure*}

\begin{figure*}
    \centering
    \includegraphics[width=\textwidth]{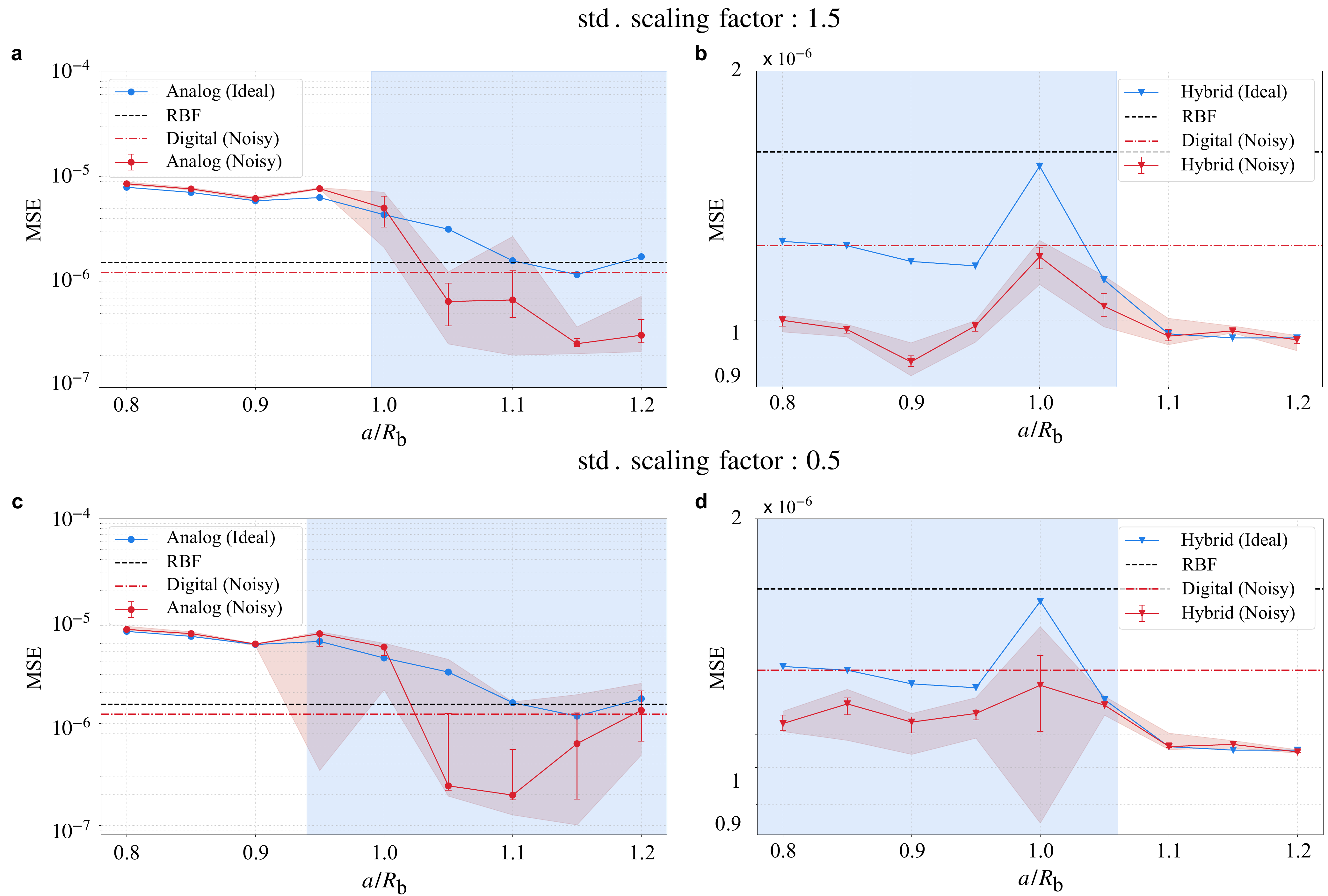}
    \caption{
    \textbf{Enhanced performance under different distributions of the noisy parameters.} We examine the performance of the noisy analog and hybrid quantum models under different specified distributions of the noise parameters. Error bars are estimated from 3 independent simulations, with the center markers indicating the median and the bars representing the first and third quartiles. The red-shaded region represents the full range of the data, whereas the blue-shaded region indicates the region where the performance enhancement has been observed. Even with different scaling factors for the standard deviations of noisy parameters, the noisy quantum model can still outperform its ideal counterpart over a specific range of interatomic distances. In the analog quantum model, this range becomes $a\geq 0.95~R_\mathrm{b}$ and $a\geq R_\mathrm{b}$ when the scaling on the standard deviation is $0.5$ and $1.5$, respectively. In contrast, the range is $a\leq1.05~R_\mathrm{b}$ for the hybrid quantum model, which is consistent with the range under the distributions in the main text.
    }
    \label{fig: different gaussian}
\end{figure*}

Here, we provide an analysis of the exponential concentration and sensitivity for the chosen distribution in the analog and hybrid quantum kernels. Analogous to the barren plateau in variational quantum circuits~\cite{Cerezo2021}, the quantum kernel exhibits exponential concentration~\cite{Thanasilp_2024}, making its experimental estimation costly in terms of the shot resources. A direct diagnostic of this concentration is the variance of the kernel matrix elements $\mathrm{Var}[\tilde{\kappa}(\vec{x}_k,\vec{x}_j)]$, and a smaller variance indicates that $\tilde{\kappa}(\vec{x}_k,\vec{x}_j)$ is clustered around the mean, so that distinguishing two kernel matrix elements requires higher precision and more shots. The variance $\mathrm{Var}[\tilde{\kappa}(\vec{x}_k,\vec{x}_j)]$ for the analog and hybrid quantum model is shown in Fig.~\ref{fig: kernel variance}. We observe that the analog quantum kernel exhibits a concentration roughly an order of magnitude stronger than that of the hybrid quantum kernel, and the presence of noise further reduces the variance of both kernels. In particular, the variance of the hybrid quantum kernel is more susceptible to the noise than that of the analog quantum model.

To illustrate the precision required for kernel estimation, we numerically round each kernel matrix element to 4, 6, and 8 decimal places and retrain the associated quantum models. We emphasize that this does not reproduce the actual experimental measurements but only provides an approximate simulation of them. The resulting performance is shown in Fig.~\ref{fig: performance under shotnoise}. Consistent with the kernel variance, we observe that the analog quantum model requires a precision $\approx 10^{-8}$ to reproduce the performance and behavior under an exact estimation, whereas the hybrid quantum model already approaches its exact-estimation performance at a precision of $\approx 10^{-4}$. 

We further examine the performance of the noisy analog and hybrid quantum models as the device becomes either noisier (i.e., with a larger variance in the noise parameters) or closer to ideal (i.e., with a smaller variance). Specifically, we scale the standard deviation of the noise parameters by factors of $1.5$ and $0.5$ and reconstruct the noisy quantum models under these settings, with each configuration simulated three times. The numerical results are shown in Fig.~\ref{fig: different gaussian}. Even with different scaling factors applied to the standard deviations, performance enhancement is still consistently observed, although the corresponding range of interatomic distances deviates from those reported in the main text. For the analog quantum model, the range where performance enhancement occurs shifts to $a\geq 0.95~R_\mathrm{b}$ and $a\geq R_\mathrm{b}$ for the scaling factors of $0.5$ and $1.5$, respectively. In contrast, for hybrid quantum model, this range remains largely consistent with the range $a\leq1.05~R_\mathrm{b}$ as reported in the main text. This demonstrates that the presence of noise can be counterintuitively beneficial to the analog and hybrid quantum models under different distributions of the noise parameters.

\thispagestyle{empty}
\bibliography{ref}
\end{document}